\documentclass[aps, pra, reprint, amsmath,amssymb,graphicx]{revtex4-1}

\usepackage{mathtools}
\usepackage{amssymb}
\usepackage{amsfonts}
\usepackage{amsmath}
\usepackage{amsthm}
\usepackage{dsfont}

\usepackage{braket}

\usepackage{qcircuit}

\usepackage{graphicx}
\usepackage{xcolor}

\usepackage{hyperref}
\usepackage[capitalise]{cleveref}

\crefname{pluralequation}{eqs.}{eqs.}

\newcommand{\leftlabel}[1]{\push{\phantom{#1}}& \lstick{#1}}

\newcommand{\ketbra}[2]{| #1\rangle\! \langle #2|}
\newcommand{\M}[2][]{\mathcal{M}_{#1}\left( #2 \right)}

\newcommand{\eg}{e.\,g.\ }
\newcommand{\ie}{i.\,e.\ }
\newcommand{\etal}{\emph{et al.\ }}

\DeclareMathOperator*{\Tr}{Tr}
\DeclareMathOperator*{\Var}{Var}
\DeclareMathOperator*{\erf}{erf}
\DeclareMathOperator*{\atan2}{atan2}

\theoremstyle{plain}
\newtheorem{thm}{Theorem}[section]
\newtheorem{lem}[thm]{lemma}
\theoremstyle{definition}

\begin{document}
\title{Generating Grid States From Schr\"odinger Cat States without Post-Selection}
\author{Daniel J. Weigand}
\affiliation{QuTech, Delft University of Technology, Lorentzweg 1, 2628 CJ Delft, The Netherlands}
\author{Barbara M. Terhal}
\affiliation{QuTech, Delft University of Technology, Lorentzweg 1, 2628 CJ Delft, The Netherlands}
\affiliation{Forschungszentrum J\"ulich GmbH, J\"ulich, Germany}
\date{\today}

%
%
%
%

\begin{abstract}
Grid (or comb) states are an interesting class of bosonic states introduced by Gottesman, Kitaev and Preskill \cite{Gottesman.etal.2001:GKPcode} to encode a qubit into an oscillator.
A method to generate or `breed' a grid state from Schr\"odinger cat states using beam splitters and homodyne measurements is known \cite{Vasconcelos.etal.2010:GKPprep}, but this method requires post-selection.
In this paper we show how post-processing of the measurement data can be used to entirely remove the need for post-selection, making the scheme much more viable.
We bound the asymptotic behavior of the breeding procedure and demonstrate the efficacy of the method numerically.
 \end{abstract}
\maketitle

\section{Introduction}

Grid (or comb) states are a class of bosonic states with various interesting possible applications.
Grid states were introduced in \cite{Gottesman.etal.2001:GKPcode} as simultaneous eigenstates of two commuting displacement operators.
In this scheme grid states can be used to encode a qubit (or qudit) into an oscillator or bosonic mode so that small displacement errors can be corrected.
As outlined in \cite{Gottesman.etal.2001:GKPcode}, universal quantum computation can be achieved using grid states:
Clifford gates can be implemented via linear optics while one may invoke magic-state-distillation techniques to get to universality.
Grid states also play a crucial role in fault-tolerant continuous-variable computation using cluster states \cite{Menicucci.2014:ClusterStateComp}.

It has also been shown that grid states can be used to generate maximal violations of CHSH-type inequalities \cite{WHG03:Grid_state-Bell,Etesse.etal.2014:GridBellTest}. In recent work, we have shown that a grid state can be used to determine the two parameters of a small displacement accurately and simultaneously \cite{Duivenvoorden.etal.2017:Sensorstate}, going beyond squeezed or coherent states. 


First proposals to generate grid states use, \eg, the coupling between a micro-mirror and an optical mode \cite{Gottesman.etal.2001:GKPcode}, the oscillatory motion of a trapped atom \cite{Pirandola.etal.2006:GKPprepNeutrAtoms,Travaglione.Milburn.2002:GKPprepStandardPE} or a Kerr interaction between two bosonic modes \cite{Pirandola.etal.2004:GKPprepKerr}.
Recent ideas on generated grid states in an atomic ensemble using squeezed light can be found in \cite{Motes.etal.2017:QuantumPhysics}, while an optical breeding protocol for cat states was considered in \cite{Sychev.etal.2017:CatBreeding}. In earlier work, we have shown how grid states can be generated without post-selection using phase estimation and a qubit-bosonic mode coupling of the form $Z a^{\dagger}a$ \cite{Terhal.Weigand.2016:GKPprepPE}, focusing on a circuit-QED setting. Very recent experiments \cite{fluehmann+home,kienzler+:iontrap} show how a grid state can be constructed in the motional mode of an ion using post-selection.

In the linear optics setting, Vasconcelos \etal \cite{Vasconcelos.etal.2010:GKPprep} and Etesse \etal \cite{Etesse.etal.2014:GridBellTest} have independently developed a \emph{breeding} protocol to generate grid states from Schr\"odinger cat states, using linear optics and homodyne post-selection \cite{Vasconcelos.etal.2010:GKPprep}.
A similar breeding protocol, used to generate Schr\"odinger cat states from Fock states, has been demonstrated in an experiment by Etesse \etal \cite{Etesse.etal.2015:CatBreedingExp}.
However, the protocol has an important drawback:
The success probability of post-selection diminishes rapidly with the number of rounds.

In this paper, we show that classical post-processing can be used to correct the grid state generated by a breeding protocol. This allows the use of any state generated by breeding, independent of the measurement results, showing that no post-selection is necessary. 
Our understanding of the protocol is formed by showing that a breeding protocol has identical action as a phase estimation protocol of multiple rounds, with specific (known) feedback phases and measurement results.
Through this identification the breeding protocol implements a particular phase estimation protocol which by definition gradually projects onto a grid state (since one is gradually learning bits of the phase).
The feedback phases used and bits obtained in phase estimation inform us about the grid state that we have obtained, namely the information gives us an estimate of the eigenvalues of the commuting displacement operators thus fixing the eigenstate.

By describing a toy model, the so-called slow breeding protocol, we can show how breeding can be related to phase estimation. However, this slow breeding protocol is non-optimal in its requirement for very large cat states. We then examine an efficient breeding protocol, which is the protocol in \cite{Vasconcelos.etal.2010:GKPprep}, and show how the measurement record can be used to correct any final state to a good grid state. 
Proving convergence of this breeding protocol towards a good grid state by invoking phase estimation is not simple.
Instead, by using a new class of approximate grid states which is closed under the efficient breeding step, we can bound the asymptotic behavior of the protocol. Finally, we confirm the performance of the protocol with numerics.


We will first review some background concepts concerning grid states, squeezing parameters and phase estimation in \cref{sec:background}.
In \cref{sec:breeding} we show how a breeding protocol can be mapped onto a phase estimation scheme, giving some intuition how a protocol works without post-selection.
Then we focus on analyzing the efficient breeding protocol by Vasconcelos \etal \cite{Vasconcelos.etal.2010:GKPprep} without post-selection.
In \cref{sec:bounds} we introduce a very useful class of approximate grid states and present some bounds on the probability of improving the state in a breeding round using these approximate states.
We close the paper with a numerical simulation of the breeding 
protocol in \cref{sec:simulation} and a Discussion (\cref{sec:discussion}).

\section{Background}
In this section, we give a short review of previous results and the formalism needed in the rest of this paper. We start in \cref{sec:grid} with a short introduction of grid states, following mostly the paper by Gottesman \etal \cite{Gottesman.etal.2001:GKPcode}. In \ref{sec:eff_squeezing}, we review the effective squeezing parameters, a versatile metric for the quality of a grid states which we introduced in \cite{Duivenvoorden.etal.2017:Sensorstate}. In \cref{sec:ape}, we introduce a formalism which enables the construction of a map between breeding and phase estimation in an efficient manner.


\label{sec:background}
\subsection{Grid states}
\label{sec:grid}
Consider a bosonic mode with dimensionless quadrature operators $\hat{q}=\frac{1}{\sqrt{2}}(a+a^{\dagger})$ and $\hat{p}=\frac{i}{\sqrt{2}}(a^{\dagger}-a)$ obeying $[\hat{q},\hat{p}]=i$.
A grid state in this mode is a simultaneous, approximate, $+1$ eigenstate of two commuting displacement operators $S_p=e^{i u \hat{p}}$ and $S_q=e^{i v \hat{q}}$ where $u\cdot v\mod 2\pi = 0$ ensures commutativity of $S_p$ and $S_q$.
Note that it is not necessary that the displacements $S_p, S_q$ form a square lattice in phase space. In fact, grid states can be defined on any two dimensional lattice where the area of the unit cell is a multiple of $2\pi$ \cite{Gottesman.etal.2001:GKPcode}. 

In this paper, we will investigate grid states with a symmetric choice $u=v=\xi$. For example, for the choice $\xi=\sqrt{2\pi}$, the space fixed by $S_p=+1, S_q=+1$ is one-dimensional. This state will be referred to as the \emph{sensor state}\cite{Duivenvoorden.etal.2017:Sensorstate}. 


Whenever a choice for $\xi$ is necessary (\eg for the numerical analysis or the Wigner function of a state), we investigate protocols generating this sensor state. In case of the choice $\xi=2\sqrt{\pi}$ the $+1$ eigenspace of $S_p$ and $S_q$ is two-dimensional and thus encodes a qubit \cite{Gottesman.etal.2001:GKPcode}.
From here on, we will refer to $\xi$ as the {\em spacing} of a grid state.  For both the wavefunction in quadrature space and the Wigner function of a grid state, the spacing corresponds to the distance between the sharp peaks in these functions. We use the notation for displacement $D(\alpha)=\exp(\alpha a^{\dagger}-\alpha^* a)$ so that $S_p=D(\sqrt{\pi})$ for the sensor state. Spacing $\xi$ thus corresponds to the action of a displacement with coherent amplitude $\xi/\sqrt{2}$.


\begin{figure}
	\hspace{0.25cm} a \hspace{1.202cm} b \hspace{1.202cm} c \hspace{1.202cm} d \hspace{1.202cm} e
	\includegraphics[width = \hsize]{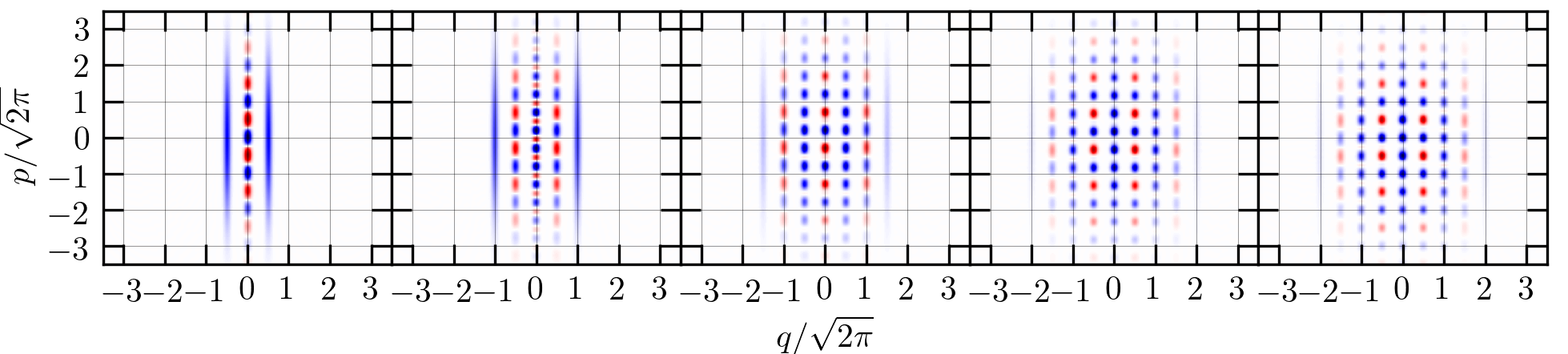}
	\includegraphics[width = \hsize]{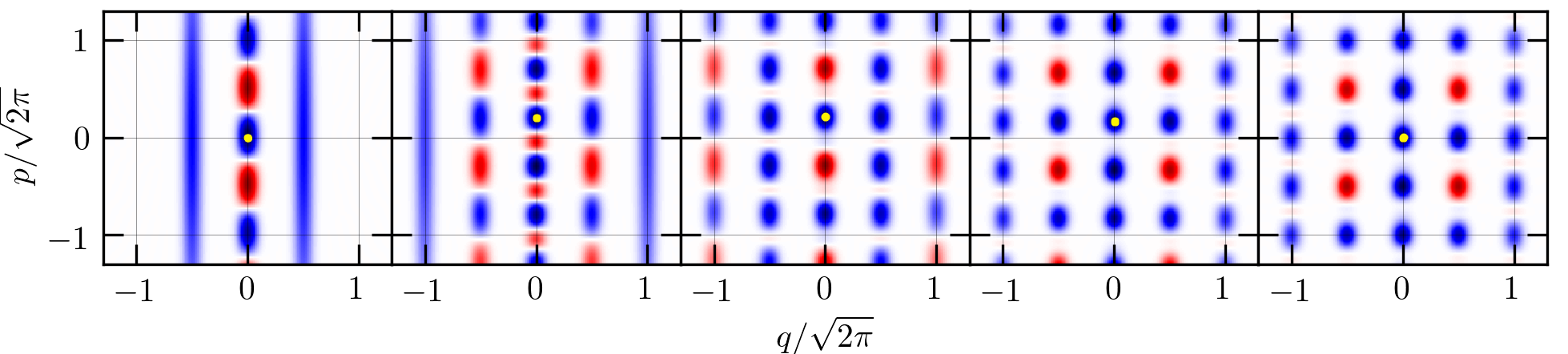}
\caption{An example of breeding of the approximate $+1$ eigenstate of $S_p = e^{i\sqrt{2\pi}\hat{p}}$ and $S_q = e^{i\sqrt{2\pi}\hat{q}}$ (sensor state). Top row: Panels (a) to (d) show the Wigner functions of states generated by resp. $N=1,2, 3,4$ measurement operators ${\cal M}$ (as in \cref{eq:PE_form} with some particular choice of phases $\varphi_j$ which provide a good illustration of how the grid is shifted) where the horizontal axis is the $q$-coordinate and the vertical axis is the $p$-coordinate. Panel (a) is the Wigner function of a squeezed cat state with squeezing parameter $\Delta=0.2$. The grid state is gradually built by displacements (translations) to the left and right with $S_p^{-1/2}$ and $S_p^{1/2}$. For the state with $N=4$ in panel (d), we show the same state after applying the correction $D_{\rm correct}$ in panel (e). Bottom row: Shown are the same Wigner functions, zoomed in around the origin. The yellow dots mark the `center' of the state, for a $+1$ eigenstate of $S_p$ it lies at the origin.}
\label{fig:grid_state}
\end{figure}



Since a perfect eigenstate of these displacement operators, \ie an ideal grid state has infinite energy, it is only possible to generate approximate grid states.
One possible approximation is a grid state of the form
\begin{align}
\ket{\Psi} \propto \sum_{t=-\infty}^\infty e^{-\pi\kappa^2 t^2}S_p^t S(\Delta)\ket{\mathrm{vac}},
\label{eq:grid}
\end{align}
where $S_p^t$ corresponds to the displacement $D(t\xi/2)$ and $S(\Delta)$ is the squeezing operator which has the action $\hat{q}\to\hat{q}\Delta, \hat{p}\to\hat{p}/\Delta$ (so that $\bra{{\rm vac}}S^{\dagger}(\Delta){\rm Var}(q) S(\Delta)\ket{{\rm vac}}=\Delta^2 \bra{{\rm vac}} {\rm Var}(q)\ket{{\rm vac}}=\frac{\Delta^2}{2}$).
The squeezing parameter $\Delta < 1$ and the width of the Gaussian envelope can be chosen to be the same, \ie $\kappa = \Delta$ \cite{Gottesman.etal.2001:GKPcode}. 

In this form, the squeezed vacuum can be understood as an approximate $+1$ eigenstate of $S_q$, while the weighed sum over powers of $S_p$ is an approximation of the projector onto the $+1$ eigenspace of $S_p$.
Essentially, the ideal grid state is invariant under the two translations $S_p$ and $S_q$ (and their inverses) in phase space,  hence a $+1$ eigenstate of these operators.
Any finite-photon number version of this state occupies a bounded volume in phase space and cannot be fully translationally-invariant, but a Gaussian envelope allows the non-translational invariance of the tails to play a relatively small role.

\subsection{Effective squeezing parameters}
\label{sec:eff_squeezing}
In order to characterize the quality of an approximate grid state we have introduced so-called effective squeezing parameters for both quadratures in \cite{Duivenvoorden.etal.2017:Sensorstate}.
A `squeezing' parameter can be generally used for capturing how well a state $\rho$ is an approximate eigenstate of a unitary operator $U$.
The idea is based on the fact that a state $\rho$ is an eigenstate of the operator $U$ iff $|\Tr \rho U| = 1$.
For such a state the mean phase $\theta \in [-\pi,\pi)$ equals $\theta (\rho)= \arg(\Tr U \rho)$.
Because of the $2\pi$-periodicity of the phase, the variance should not be taken to be the standard variance, but can be chosen as a phase variance equal to $\Var(\rho) = \ln(|\Tr U \rho |^{-2})$ \cite{Duivenvoorden.etal.2017:Sensorstate}.
This variance is identical to the more commonly used Holevo phase variance \cite{book:WM} for small $|\Tr U \rho |$. 

For a displacement $\mathcal{D}:=D(u e^{i\phi})$ with $\phi, u\in\mathds{R}$, the variance should be rescaled by $u$, \ie we define the mean phase $\theta_\mathcal{D}$ and the effective squeezing parameter $\Delta_\mathcal{D}$ as:
\begin{align}
&\theta_\mathcal{D} := \arg (\Tr \mathcal{D} \rho), &&\Delta_\mathcal{D} := \frac{1}{u}\sqrt{\ln(|\Tr \mathcal{D} \rho|^{-2})}.
\label{eq:delta}
\end{align}
As grid states are defined with respect to the displacement $S_p$ ($S_q$) along the real (imaginary) axis in phase space, it is convenient to use the short-hand $\Delta_p:=\Delta_{S_p}$ and $\Delta_q:=\Delta_{S_q}$ for the two \emph{effective squeezing parameters}.
The squeezing parameters of an approximate grid state as defined in \cref{eq:grid} are $\Delta_q=\Delta$, $\Delta_p \approx \kappa$.
For a squeezed vacuum state $S(\Delta) \ket{\rm vac}$, one has $\Delta_q=\Delta=1/\Delta_p$. 
The effective squeezing parameter and mean phase have a very natural relation to grid states:

Protocols to generate an approximate eigenstate of $S_p$ and $S_q$ will produce a state $\rho$ with certain values for $\theta_p:= \theta_{S_p}$, $\theta_q:=\theta_{S_q}$, $\Delta_p$ and $\Delta_q$. The effective squeezing parameters then give a direct measure of the quality of the state $\rho$. In case of the sensor state, they directly relate to the measurement precision that can be achieved using $\rho$ as a sensor \cite{Duivenvoorden.etal.2017:Sensorstate}. In case of the GKP code, the probability of a logical X (or Z) error in the encoding can be bounded as $P_{\text{error}} < \frac{2\Delta}{\pi}e^{-\pi/(4\Delta^2)}$ with $\Delta=\Delta_q=\Delta_p$ \cite{Gottesman.etal.2001:GKPcode}.

The mean values $\theta_p$ and $\theta_q$ which are extracted from the protocol can be used to correct the resulting state by displacing this state by $D_{\rm correct}$, i.e. $\rho \rightarrow \rho'=D_{\rm correct} \rho D_{\rm correct}^{\dagger}$ such that $\theta_p(\rho')\approx \theta_q(\rho') \approx 0$. For example, to shift the mean phase $\theta_p$ back to 0 we choose $\alpha$ in $D_{\rm correct}=\exp(i \alpha \hat{q})$ such that $S_p D_{\rm correct}=\exp(-i \theta_p) D_{\rm correct} S_p$. A simple visual representation of this procedure is that the positive parts of the Wigner function form a grid in phase space for grid states and this grid is aligned with the $p=0,q=0$ axes for a $+1$ eigenstate of $S_p,S_q$ (see \cref{fig:setup,fig:grid_state}).

The final state is then an approximate $+1$ eigenstate of $S_p$ and $S_q$.
However, it is not necessary to perform such a correcting displacement if one uses the concept of a phase or displacement frame \cite{Terhal.Weigand.2016:GKPprepPE} (in analogy with a Pauli frame for qubits).

Clearly, approximate grid states are not unique. For example, two grid states whose grid envelope is displaced or translated one unit cell over can have the same values for $\theta_p, \theta_q$ and $\Delta_p,\Delta_q$ but contain a different mean number of photons.
Similarly, one can note that the corrective displacement is not unique:
in practice one may opt for the smallest displacement shifting the grid envelope to the correct position, see \cref{fig:grid_state}(e).


\subsection{Adaptive phase estimation}
\label{sec:ape}
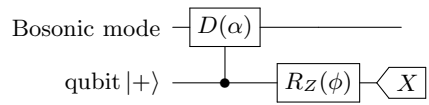
\begin{figure}
\begin{minipage}{.3\textwidth}
	\Qcircuit @C=.7em @R=.7em {
		\leftlabel{\mbox{Bosonic mode}} & \gate{D(\alpha)} & \qw & \qw \\
		\leftlabel{\mbox{qubit} \ket{+}} & \ctrl{-1} &  \gate{R_Z(\phi)} &\measuretab{X}
	}
\end{minipage}
\caption{Single round of an adaptive phase estimation protocol which estimates the eigenvalue of $D(\alpha)$. The output state goes back into the next round of the protocol and the feedback phase $\phi$ can be chosen depending on earlier rounds. The collection of bits obtained, together with the chosen feedback phases, will gradually project the input state onto an approximate eigenstate of $D(\alpha)$ as the approximate eigenvalue is learned.}
\label{fig:PE_round}
\end{figure}

Phase estimation refers to a whole class of algorithms that measure the eigenvalue of a unitary operator $U$.
A recent overview of some of these schemes can be found e.\,g.\ in \cite{Svore.etal.2014:PhaseEstimation}.
All phase estimation procedures, including textbook phase estimation \cite{book:Nielsen.Chuang.2000:QCompAndQInfo}, Kitaev's phase estimation \cite{Kitaev.1995:RPE} and variants thereof, can be executed in an iterative form with a single-qubit applying controlled-$U^k$ gates.
An in-depth analysis of some adaptive schemes can be found in works by Berry \etal \cite{Berry.etal.2001:ARPE}. We are interested in this case when the unitary operator to be measured is some displacement and we consider performing such measurement by repeating a circuit of the form \cref{fig:breeding}.

A convenient formalism to describe such adaptive phase estimation uses the following `measurement' operator:
\begin{align}
\M{\varphi, \alpha} := 1 + e^{i\varphi}D\left(\alpha\right),
\label{eq:M}
\end{align}
with $\varphi\in [0, 2\pi)$ and $\alpha$ is a coherent amplitude.
 In all what follows we focus on breeding an approximate eigenstate of $S_p=D(\xi/\sqrt{2})$ and assume that $\alpha$ is real, but the same method can be used for complex $\alpha$.
 
 With this operator, a squeezed Schr\"odinger cat state has the form  $D(-\alpha/2)\M{0,\alpha}S(\Delta)\ket{\text{vac}}$, \ie a single application of the measurement operator onto a squeezed vacuum state, plus an additional displacement.
  
One can also see that the circuit shown in \cref{fig:PE_round} acts on an input state $\ket{\Phi_0}$ as $\M{\phi+\pi x, \alpha}\ket{\Phi_0}$ where $x\in\{0,1\}$ is the measurement result, \ie it applies one additional measurement operator to the initial state.
Thus \emph{any} state generated by a sequence of $N$ of these circuits is of the form
\begin{align}
\ket{\Psi} \propto \prod_{j=1}^N \M{\varphi_j, \alpha_j}\ket{\Phi_0},
\label{eq:PE_form}
\end{align}
where $\ket{\Phi_0}$ is the initial state and $\varphi_j=\phi_j+x_j \alpha_j$ with measurement outcome $x_j$, feedback phase $\phi_j$ of round $j$ and 
$\alpha_j$ possibly varying per round. 

It can be observed that the class of states which is described by fixing the outcome to be $x=0$ and letting the feedback phase vary captures all states in \cref{eq:PE_form} since $\phi_j\in[0,2\pi)$ can be freely chosen.
We will show that the state obtained by a breeding protocol is identical to a state obtained by such a phase estimation protocol with all outcomes $x_j=0$ and with varying $\phi_j=\varphi_j$.
 A (trivial) example is that a squeezed Schr\"odinger cat state is equivalent to a single round of phase estimation with $x=\phi=0$ applied to a squeezed vacuum state.
This map gives some intuition why breeding gives rise to a grid state. Using the form of the state allows one to estimate the mean phase and the effective squeezing parameters of the state using \cref{eq:delta}.

As was mentioned before, even given $\theta_p$ and $\Delta_p$, a grid state is not unique since it can be shifted by any $S_p$ without affecting these parameters.
Thus to place the grid state symmetrically around the vacuum state and minimize photon number, it is better to perform a pre-displacement by $D(-\alpha/2)$ in each phase estimation round in Fig.~\ref{fig:PE_round} and similarly use the measurement operator $D(-\alpha/2)+ e^{i\varphi} D(\alpha/2)$.
Since our analysis does not depend on these shifts, we have opted to not include them.

\section{Breeding}
\label{sec:breeding}
\begin{figure}
\begin{minipage}{.15\textwidth}
	\Qcircuit @C=.7em @R=.7em {
		\leftlabel{\prod_{j=1}^{N_1} \M{\varphi_j,\alpha_j}\ket{\Phi_0}} & \multigate{1}{\mbox{50:50 BS}} & \qw & \push{\phantom{\hspace{3cm}}}&\lstick{\prod_{j=1}^{N_1+N_2} \M{\tilde{\varphi}_j,\tilde{\alpha}_j}\ket{\Phi_0}}\\
		\leftlabel{\prod_{j=1}^{N_2} \M{\psi_j,\beta_j}\ket{\Phi_0}} & \ghost{\mbox{50:50 BS}} &  \measureD{p}
	}
\end{minipage}
\caption{Single round of a breeding protocol. The round takes approximate grid states defined by $N_1$ and $N_2$ measurement operators $\mathcal{M}$ as inputs. It returns an approximate grid state with $N_1+N_2$ (possibly changed) measurement operators. The second output port is subject to a homodyne measurement of the $\hat{p}$ quadrature. The initial state $\ket{\Phi_0}$ is chosen to be invariant under the action of a beam splitter, \eg a squeezed vacuum state.}
\label{fig:breeding}
\end{figure}
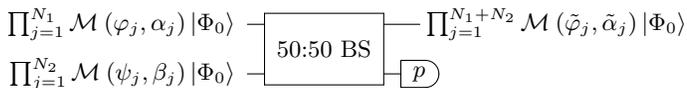
Breeding protocols refer to a procedure where a grid state is gradually constructed from input (squeezed) Schr\"odinger cat states.
One can view these input states as a very poor approximation (panel (a) in \cref{fig:grid_state}) to a grid state and the goal is to gradually improve these states.
The circuit in \cref{fig:breeding} shows a single round of breeding.
We will denote the number of breeding rounds by $M$, while $N$, which is a function of $M$, refers to the number of measurement operators acting on some initial state as in \cref{eq:PE_form}.
In a single breeding round partially bred grid states that will be of the same form as \cref{eq:PE_form} are fed into a beam-splitter.
After the beam-splitter, the $p$-quadrature of one of the states is measured (for breeding of a $S_p$ eigenstate). For $N_2=1$ in \cref{fig:breeding}, the input of the bottom port (port 2) plays the role of squeezed cat state modulo the additional pre-displacement, i.e. $(D(-\beta/2) + D(\beta/2))\ket{\Phi_0}=D(-\beta/2) \M{0, \beta}\ket{\Phi_0}$.
The aim of the Breed operation is to map the measurement operators in port $2$ to port $1$, \ie the state at the output port is still of the form of \cref{eq:PE_form}, but with $N_1+N_2$ measurement operators.

Since we would like to produce a state which is both an approximate eigenstate of $S_p$ and $S_q$, we choose the input state $\ket{\Phi_0}$ as a squeezed vacuum state $\ket{\Phi_0}=S(\Delta) \ket{\rm vac.}$ providing an approximate eigenstate of $S_q$.
 It is important that the effective squeezing parameter $\Delta_q$ is approximately preserved under the breeding operation so that the outgoing state is both an approximate eigenstate of $S_p$ and $S_q$: we will verify this at the end of Section \ref{sec:eff_breeding}.
 
The rounds of this breeding procedure could be repeated in at least two ways.
In the first manner, which we call slow breeding, we always use a squeezed cat state at input port 2 and input port 1 contains the state that came out of port 1 in the previous breeding round.
This protocol can be seen as a toy model in that it has several drawbacks, but we describe its functionality in order to understand how breeding works and how it maps onto phase estimation.
In Section \ref{sec:eff_breeding} we describe a parallelized distillation protocol in which $2^M$ squeezed cat states are fed into beam-splitters, leading to $2^{M-1}$ output states, which are subsequently used to produce $2^{M-2}$ states etc., eventually extracting one grid state after $M$ breeding rounds, \ie the setting proposed in \cite{Etesse.etal.2014:GridBellTest,Vasconcelos.etal.2010:GKPprep}. Then, we will show how a map to phase estimation can be constructed for this protocol, removing the need for post-selection.

\subsection{Slow Breeding}
Using that the action $\mathcal{B}$ of the beam splitter is given by
\begin{align*}
&\hat{q}_1\to (\hat{q}_1 - \hat{q}_2)/\sqrt{2}, && \hat{p}_1\to (\hat{p}_1 - \hat{p}_2)/\sqrt{2},\\
&\hat{q}_2 \to (\hat{q}_1 + \hat{q}_2)/\sqrt{2}, && \hat{p}_2\to (\hat{p}_1 + \hat{p}_2)/\sqrt{2},
\end{align*}
one can show that the output state of a breeding round in \cref{fig:breeding} equals 
\begin{align}
\prod_{j=1}^{N_1} \prod_{k=1}^{N_2} \mathcal{\tilde{M}}_1(\varphi_j,\alpha_j)
\mathcal{\tilde{M}}_2(\psi_k, \beta_k)\mathcal{B} \ket{\Phi_0, \Phi_0}.
\label{eq:BSO_output}
\end{align}
Here $\mathcal{\tilde{M}}_i(\varphi,\alpha)=\mathcal{B} \mathcal{M}_i(\varphi,\alpha) \mathcal{B}^{\dagger}$. For the input states $\ket{\Phi_0,\Phi_0}$ we use the invariance under beam-splitting, i.e. $\mathcal{B} S_{1}(\Delta)S_2(\Delta) \ket{\mathrm{vac, vac}}_{1,2} = S_{1}(\Delta)S_2(\Delta) \ket{\mathrm{vac, vac}}_{1,2}$.
For real $\alpha$ we have $\mathcal{\tilde{M}}_1(\varphi,\alpha)=I+e^{i\varphi} D_1(\alpha/\sqrt{2}) D_2(-\alpha/\sqrt{2})$ and 
$\mathcal{\tilde{M}}_2(\psi,\beta)=I+e^{i\psi}D_1(\beta/\sqrt{2}) D_2(\beta/\sqrt{2})$.
When mode 2 is then measured via homodyne measurement of $\hat{p}$ with outcome $p$, we can replace $D_2(\alpha)$ by $e^{-i \alpha \sqrt{2} p}$ (for real $\alpha$).
This implies that the output state of the protocol is as claimed in Fig.~\ref{fig:breeding}, \ie
\begin{align}
&\prod_{j=1}^{N_1}\prod_{k=1}^{N_2} \mathcal{M}_1(\tilde{\varphi}_j, \tilde{\alpha}_j) \mathcal{M}_1(\tilde{\psi}_k, \tilde{\beta}_k) \ket{\Phi_0}, \label{eq:bso} \\
&\tilde{\varphi}_{j}=\varphi_j+ \alpha_j p,\ \tilde{\alpha}_{j}=\frac{\alpha_j}{\sqrt{2}},\  \tilde{\psi}_{k}=\psi_k-\beta_{k} p,\ \tilde{\beta}_{k}=\frac{\beta_k}{\sqrt{2}}. \notag
\end{align}
The probability to find outcome $p$ for the homodyne measurement depends in detail on the state of the form \cref{eq:PE_form} that goes into the beam-splitter, but the variance of this probability distribution in $p$ scales as $\sim 1/\Delta^2$. Hence the more the input state $\ket{\Phi_0}$ is squeezed in $q$ (by $\Delta$), the large the spread of measured values for $p$ will be and hence the greater the need for not using post-selection on the outcome $p=0$.

Consider now the {\em slow breeding} case where the state at input $2$ is always a squeezed cat state \ie $N_2=1$ , and the output state is fed into port $1$ of the next round.
In order to breed a grid state we take $\alpha_1=\beta_1=\alpha$ and $\varphi_1=\psi_1=0$ for the first breeding round, meaning that the inputs in both ports are squeezed cat states.

In the second breeding round one takes $\beta_2=\alpha/\sqrt{2}, \psi_2=0$ and in the $M$th round $\beta_M=\alpha/\sqrt{2^{M-1}}, \psi_M=0$ so that the final state has spacing $\xi=\alpha/\sqrt{2^{M-1}}$. The evolution of mode 1 under the slow breeding protocol without post-selection and $M=3$ rounds is shown in \cref{fig:grid_state}(a-d).

By post-selecting the measurement result onto $p=0$, it is apparent from this choice for the $\beta_i$ and \cref{eq:bso} that $M$ rounds of this procedure generate a binomial distribution of displacements, since all the phases are zero.
 Thus, clearly, when we post-select on outcome $p=0$, one obtains a grid state with a binomial envelope (similar to the protocols shown in \cite{Vasconcelos.etal.2010:GKPprep,Etesse.etal.2014:GridBellTest}). 

From \cref{eq:bso} it follows immediately that $M$ rounds of breeding in this setup with a final spacing $\xi=\alpha/\sqrt{2^{M-1}}$ can be mapped to $N=M+1$ rounds of phase estimation with the choice $\varphi_m = \alpha (\sum_{k>m}^M 2^{-k/2} p_k - 2^{-m/2} p_m)$ for the feedback phase and measurement result $x_m=0$, where $p_m$ is the homodyne measurement result of $\hat{p}_2$ in round $m=1, \ldots, M$ and $p_0=0$ to fix the initial state ($m=0$) to the squeezed cat state $\propto (I + D(\alpha))\ket{\Phi_0}$.
It is noteworthy that the feedback phase depends on the outcomes of many `later' rounds: One can thus only construct the corresponding phase estimation protocol after the last homodyne measurement is done.
This suggests that instead of post-selecting on $p=0$, one can simply process the measurement information to infer the values of $\theta_p$ in Eq.~(\ref{eq:delta}) of the final state. This correction is demonstrated in \cref{fig:grid_state}(e), where a correcting displacement is applied to the final state of the protocol.

However, the slow breeding protocol suffers from a different problem. To get a grid state with final spacing $\xi=\sqrt{2\pi}$ after $M$ rounds, the number of photons in the squeezed cat state used in the first round $\bar{n}_{cat} \geq 2^M \pi$.
This is exponentially larger than the mean photon number of the final grid state which scales as $\bar{n}_{grid} \sim M$ \cite{Terhal.Weigand.2016:GKPprepPE,Gottesman.etal.2001:GKPcode}, \ie the procedure is inefficient in its use of photons. 

\subsection{Efficient Breeding}
\label{sec:eff_breeding}


A much better scheme is to use a partially-bred grid state in the ancilla mode as proposed in \cite{Vasconcelos.etal.2010:GKPprep,Etesse.etal.2014:GridBellTest}, effectively performing a grid state distillation scheme.
In this scheme, one starts with two cat states ($N_1=N_2=1$), leading to a state with $N_{\rm out}=2$. Then one takes two such states ($N_1=N_2=2$) and feeds them into the beam-splitter to get a state with $N_{\rm out}=4$ etc.
With \cref{eq:bso}, one can see that we have $N=2^M$ for $M$ repetitions of this scheme.

In this scheme one will always have $\beta_j=\alpha_j$ for the two input ports, but the phases can vary depending on measurement results and do not need to be the same for both inputs.
This parallelization leads to a much faster built-up of the grid state.
For a final grid state with $N=2^M$ applications of $\mathcal{M}$, one requires $M$ rounds of beam-splitters in sequence.


For the final grid state to have spacing $\xi$ one starts the protocol with cat states with amplitude $\xi 2^{(M-3)/2}=\xi\frac{\sqrt{N}}{2\sqrt{2}}$, thus $\bar{n}_{cat} \sim \bar{n}_{grid}\sim N$. For example, generating a sensor state with $M=2$ rounds would require $\bar{n}_{cat} = \frac{\pi}{2} + \bar{n}_{sq}$ photons, where $\bar{n}_{sq}$ is the additional number of photons due to squeezing.\\

\begin{figure}
	\includegraphics[width=\hsize]{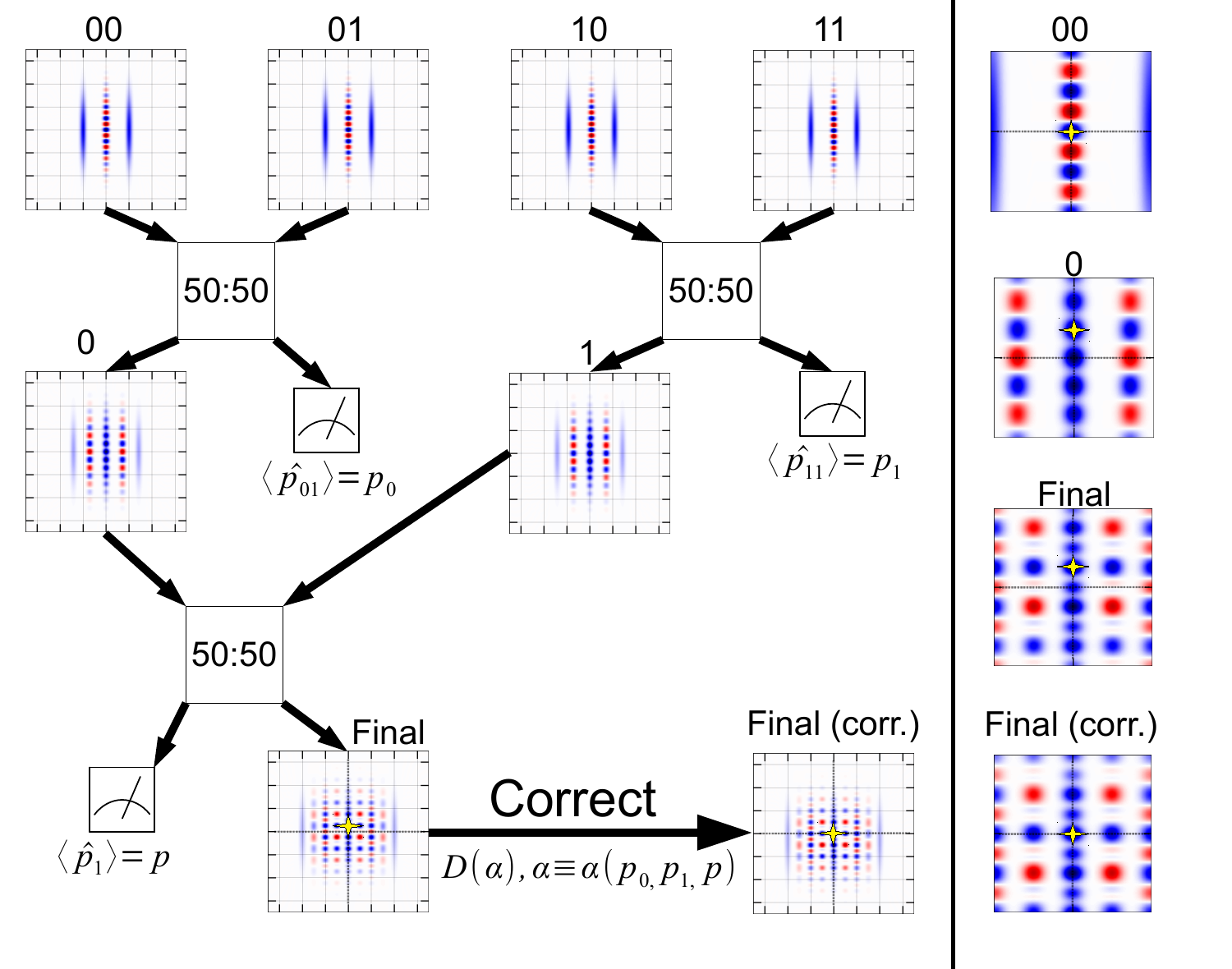}
	\caption{Left side: Efficient breeding protocol as proposed in \cite{Vasconcelos.etal.2010:GKPprep,Etesse.etal.2014:GridBellTest}, with $M=2$ rounds but without post-selection. The labeling of modes is according to the scheme introduced in \cref{sec:eff_breeding}: The initial $2^M=4$ Schr\"odinger cat states are labeled by the 2-bit strings $\{00,01,10,11\}$. Those are put pairwise into beam splitters, resulting in the states and measurement results labeled by the 1-bit strings $\{0,1\}$. The phases and final state of the corresponding phase estimation setup are determined using \cref{eq:effective_correction}, the correcting displacement is then obtained with \cref{eq:delta}.
	Right side: Shown are the Wigner functions of modes $\{00, 0, \text{Final}\}$ and the fianl state after applying the correction, zoomed in around the origin. The yellow crosses mark the `center' of the state, for a $+1$ eigenstate of $S_p$ they lie at the origin.
	}
	\label{fig:setup}
\end{figure}

In order to estimate the effective squeezing after $M$ rounds as well as the phase $\theta_p$, one needs to describe the final state in terms of the measurement outcomes.
A concise description of the output state of an $M$-round protocol is as follows. 
We label all $2^M$ ingoing modes of the protocol with a bit-string ${\bf x}[M]$ of length $M$.
Two modes $x_1\ldots x_{M-1} x_M$ and $x_1 \ldots x_{M-1} \overline{x_M}$ which differ only on the last bit $x_M$ will enter into one beam-splitter and so the outgoing single mode can be labeled by the remaining $M-1$ bit string $x_1..\ldots x_{M-1}= {\bf x}[M-1]$.
The outcomes of these $2^{M-1}$ measurements of the first round forms a vector ${\bf p}^1$ with $2^{M-1}$ entries $p^1_{{\bf x}[M-1]}$ which are labeled by the bitstrings ${\bf x}[M-1]$.
The final measurement in round $M$ is then ${\bf p}^{M}$ with a single entry labeled by a bit string of length 0. An example of this labeling can be seen in \cref{fig:setup}.
With this notation the initial state is thus a product state proportional to $\prod_{\mathbf{x}[M]} {\cal M}_{\mathbf{x}[M]}(0, 2^{(M-1)/2}\xi)\ket{\Phi_0}_{\mathbf{x}[M]}$.


Similarly, the state of the system after the first round of breeding is the product state 
\begin{align*}
\prod_{\mathbf{x}[M]} \M[{{\bf x}[M-1]}]{\xi(-1)^{x_M}2^{\frac{M-2}{2}}p^1_{{\bf x}[M-1]}, 2^{\frac{M-2}{2}}\xi} \nonumber \\ \ket{\Phi_0}_{{{\bf x}[M-1]}}, \nonumber
\end{align*}
where each state now gets two measurement operators applied to it since we are taking the product over all bit-strings of length $M$.
After all $2^M-1$ measurements, the final state is given by $2^M$ measurement operators acting on a single mode, i.e. 
\begin{align}
&\ket{\Psi_{\rm out}({\bf p}^{1},\ldots {\bf p}^{M})} \nonumber \\ &=\prod_{\mathbf{x}[M]}\M[{x[0]}]{\xi\sum_{j=1}^{M} (-1)^{x_j} 2^{\frac{j-2}{2}}p^{M-j+1}_{{\bf x}[j-1]},\frac{\xi}{\sqrt{2}}} \ket{\Phi_0}_{\mathbf{x}[0]}.
\label{eq:effective_correction}
\end{align}
with normalization $\mathds{P}({\bf p}^{1},\ldots, {\bf p}^{M})=\bra{\Psi_{\rm out}}\Psi_{\rm out}\rangle$.



In order to evaluate $\theta_p=\frac{\arg \bra{\Psi_{\rm out}} S_p \ket{\Psi_{\rm out}}}{\braket{\Psi_{\rm out}|\Psi_{\rm out}}}$ for a given series of outcomes ${\bf p}_{\rm all}\colon= {\bf p}^{1},\ldots, {\bf p}^{M}$, it is convenient to write down the initial state as a wavefunction in $p$ and use that $S_p \ket{p}=e^{i\xi p}\ket{p}$. The effect of this correction is shown in \cref{fig:setup}.
Similarly, one can evaluate the average $\langle \Delta_p\rangle =\sum_{{\bf p}_{\rm all}} \mathds{P}({\bf p}_{\rm all}) \Delta_p({\bf p}_{\rm all})$. 
Note that, if minimizing the run-time of this procedure is crucial (\eg for feedback in an experiment), the mean phases could be approximated using the mode of the probability distribution in $p$ corresponding to the final state.





While the map between breeding and phase estimation derived in the previous section suggests that $\langle \Delta_p\rangle$ will decrease rapidly with breeding rounds, it is in fact not simple to use this mapping to analytically prove this.
The difficulty is that since the phases can vary per round (depending on the homodyne measurement outcomes), arguments which use laws of large numbers, which apply when identical experiments are repeated, are not directly applicable. 

In order to understand the outgoing state in terms of the initial squeezing, we note that the final state of the breeding protocol after $M$ rounds consists of applying powers of $S_p=D(\xi/\sqrt{2})$ (with phases) to the initial state $\ket{\Phi_0}$ and $S_q$ commutes with $S_p$. However, the input state is not an {\em exact} eigenstate of $S_q$. Furthermore, the full description of the unitary evolution involves the beam-splitter and the measured ancilla modes and the full action does not commute with $S_q$. 

This means that a few steps are required to show that the expectation value of $S_q$ of the output state is close to the expectation value of $S_q$ of the input state. The output state of any breeding protocol will be $\ket{\Psi_{\rm out}}=A \ket{\Phi_0}$ and $A=\sum_{j=-\infty}^{\infty} \alpha_j S_p^j$
 (Where the number of non-zero coefficients $\alpha_j\in\mathds{C}$ is determined by the protocol).
  We can compute the normalization of $\ket{\Psi_{\rm out}}$ by writing the initial squeezed state as a wave function in $\ket{q}$ 
\begin{align*}
&\bra {\Psi_{\rm out}} \Psi_{\rm out}\rangle  \nonumber \\
& =\sum_{j,k=-\infty}^{\infty} \frac{\alpha_j \alpha_k^*}{\sqrt{\pi\Delta^2}} \iint \text{d}q \,\text{d}q'\  e^{-\frac{q^2}{2\Delta^2}} e^{-\frac{(q'-(j-k)\xi)^2}{2\Delta^2}}  \braket{q|q'} \\
&=  \sum_{j,k}\frac{\alpha_j \alpha_k^*}{\sqrt{\pi\Delta^2}} \int \text{d}q\  e^{-\frac{(q-(j-k)\xi/2)^2}{\Delta^2}} e^{-\frac{\xi^2(j-k)^2}{4\Delta^2}}  \nonumber \\
 & =\sum_{j,k}\alpha_j \alpha_k^*e^{-\frac{\xi^2(j-k)^2}{4\Delta^2}}.
\end{align*}
For small $\Delta \lesssim 0.5$, the last term vanishes for $j\neq k$ ($\xi$ is at least $\sqrt{2\pi}$), \ie $\sum_j |\alpha_j|^2  \approx \braket {\Psi_{\rm out} |\Psi_{\rm out}} =  1$. 
Using the same method one obtains
\begin{align*}
&\bra{\Psi_{\rm out}} S_q \ket{\Psi_{\rm out}}\\
&= \sum_{j,k}\frac{\alpha_j \alpha_k^*}{\sqrt{\pi\Delta^2}} \int \text{d}q\  e^{i\xi q}e^{-\frac{(q-(j-k)\xi/2)^2}{\Delta^2}} e^{-\frac{\xi^2(j-k)^2}{4\Delta^2}}, \\
&= \sum_{j,k}\frac{\alpha_j \alpha_k^*}{\sqrt{\pi\Delta^2}} e^{-\frac{\xi^2(j-k)^2}{4\Delta^2}}e^{-\frac{i(j-k)\xi^2}{2}}\int \text{d}q\  e^{i\xi q}e^{-\frac{q^2}{\Delta^2}}, \\
&= \sum_{j,k}\alpha_j \alpha_k^*e^{-\frac{\xi^2(j-k)^2}{4\Delta^2}}e^{-\frac{i(j-k)\xi^2}{2}}\bra {\Phi_0} S_q \ket{\Phi_0}\approx \bra {\Phi_0} S_q \ket{\Phi_0},
\end{align*}
where we used the normalization condition obtained before. This implies that $\Delta_q(\Psi_{\rm out}) \approx \Delta_q(\Phi_0)=\Delta$ for initial squeezing $\Delta \lesssim 0.5$ (which corresponds to large squeezing in $q$). The effective squeezing parameter of a squeezed Schr\"odinger cat state 
$\propto (D(-\sqrt{\pi}/2)+ D(\sqrt{\pi}/2))\ket{\Phi_0}$ is $\sqrt{\Delta^2 - \frac{2}{\pi}\ln(\tanh(\frac{\pi}{4\Delta^2}))}$, which differs from a squeezed vacuum state $\ket{\Phi_0}$ by $\mathcal{O}(10^{-17})$ for $\Delta=0.2$. This is also expected, as $\Delta_q = \Delta$ for a squeezed vacuum state and $\Delta_q \approx \Delta$ for an approximate grid state as defined in \cite{Gottesman.etal.2001:GKPcode}.


\section{Asymptotic Behavior}
\label{sec:bounds}
In this section we derive probability bounds for the breeding protocol showing how the effective squeezing parameter changes round-by-round. The known class of approximate grid states which are described by a perfect grid state to which a Gaussian distribution of shift errors is applied \cite{Gottesman.etal.2001:GKPcode} is not closed under a round of breeding, the same holds for squeezed Schr\"odinger cat states.
Thus, analyzing the effect of the breeding map for many rounds is a nontrivial problem when using either class of states.

In order to solve this issue, we introduce a new class of approximate grid states
which is closed under the breeding operation, enabling an analytical discussion. Since the breeding protocol changes the spacing of an approximate grid state round-by-round, the spacing of these states is round-dependent. To this end, we first define scale-dependent shifted grid states as
\begin{equation}
\ket{u, v, m} = \frac{\sqrt{s_m \xi}}{2\pi}e^{i\frac{v}{s_m \xi}\hat{p}} e^{i\frac{s_m \xi u}{2\pi}\hat{q}} \ket{\Psi_m},
\label{eq:def_shifted}
\end{equation}
where $u, v\in[-\pi,\pi)$. The parameter $s_m$ is some scale parameter that we will choose below, $\xi$ is the spacing of the final grid state and $\ket{\Psi_m}\propto \sum_{s=-\infty}^{\infty} \ket{p=s \xi s_m}$. With the choice $s_m=1,\xi=2\sqrt{\pi}$, one obtains the shifted code states introduced by Glancy and Knill in the context of the GKP code \cite{Glancy.Knill.2006:GKPfaultTolerance}: these states above can be viewed as an extension of this concept.
For any choice of $m$ and $s_m \xi$, it can be verified that the class of states $\ket{u,v,m}$ forms an orthonormal basis of the whole Hilbert space of the oscillator, i.e. $\braket{u,v,m|u',v',m} =\delta(u-u') \delta(v-v')$ and $\int_{-\pi}^{\pi} du \int_{-\pi}^{\pi} dv \ket{u,v,m} \bra{u,v,m}=I$, see Appendix \ref{sec:ortho}.

For our application in the breeding protocol we will choose $s_m = \sqrt{2^{m-M}}$ and one can confirm that this choice yields a shifted grid state with spacing $\xi$ for $m=M$. Note that $\ket{\Psi_m}$ is a $+1$ eigenstate of the rescaled operators $S_q^{s_m}$ and $S_p^{2\pi/(\xi^2 s_m)}$, i.e. the spacing of the states is rescaled round by round since each beam-splitter will change the spacing by $\sqrt{2}$.
We can see this by writing $\ket{\Psi_m}\propto \lim_{\Delta \rightarrow 0} \Pi_{S_{q}^{s_m}=1} S(1/\Delta) \ket{\rm vac}$ since $\lim_{\Delta\rightarrow 0} S(1/\Delta) \ket{\rm vac}=\ket{p=0}$ and  $\Pi_{S_{q}^{s_m}=1} \propto \sum_{t=-\infty}^{\infty} S_{q}^{t s_m}$ is the projector onto 
the $+1$ eigenspace of $S_q^{s_m}$.

In general, a basis of shifted grid states can be used to write down an approximate code state as a Gaussian superposition of states with different shifts  \cite{Gottesman.etal.2001:GKPcode, Terhal.Weigand.2016:GKPprepPE}. Here, we will similarly use these states but the filter for the quadrature on which we apply the breeding will not be Gaussian but determined by a von Mises probability distribution. We thus define the class of approximate shifted grid states (for general $\xi$) as
\begin{align}
\ket{V_{\kappa,\mu,m}} &:= \frac{1}{\mathcal{N}}\int_{-\pi}^\pi \text{d}u\ \int_{-\pi}^\pi  \text{d}v\  V(u-\mu)_{\kappa}\notag \\ 
&\quad\quad\times \sum_{s=-\infty}^\infty e^{i u s}G(v+2\pi s)_{s_m \Delta} \ket{u, v, m}, \label{eq:mises}\\
V(u-\mu)_{\kappa} &:= \frac{1}{\sqrt{2\pi I_0(\kappa)}}\exp\left(\frac{\kappa}{2}\cos(u-\mu)\right), \notag\\
G(v)_{\sigma} &:= \frac{1}{\sqrt{\sigma \sqrt{\pi}}}\exp\left(-\frac{v^2}{2 \sigma^2}\right). 
\label{eq:def_states}
\end{align}
In the limit of large initial squeezing $\Delta \ll 1$, the normalization constant $\mathcal{N}$ goes to 1. Note that $\mathds{P}_{\sigma}(v)=G(v)_{\sigma}^2$ is a Gaussian distribution with mean 0 and standard deviation $\sigma/\sqrt{2}$ so that when $m=M$, the standard deviation of $\mathds{P}_{\Delta s_m}(v)$ is $\Delta/\sqrt{2}$.
The choice of probability distribution on $u$ and $v$ is different because the breeding protocol acts differently on the $\hat{p}$ and $\hat{q}$ quadratures of the initial states. This choice ensures that the class of states $\ket{V_{\kappa,\mu,m}}$ is closed under breeding, see \cref{eq:breedstep}.
The probability distribution $\mathds{P}_{\kappa}(u-\mu) = V(u-\mu)^2_{\kappa}$ is the von Mises distribution and $I_\nu(\kappa)$ is the modified Bessel function of the first kind of order $\nu$.
The von Mises distribution $\mathds{P}_{\kappa}(u-\mu)$ which models a Gaussian distribution for a circular phase variable $u$ has mean $\mu$.
In the limit $\kappa \gg 1$ the probability distribution becomes Gaussian by approximating $\exp(\kappa \cos(u-\mu)) \approx \exp(-\kappa (u-\mu)^2)/2)\exp(\kappa)$ with standard deviation $1/\sqrt{\kappa}$.

The index $m=0,\dots,M$ will refer to the number of breeding rounds applied to the initial state, with $m=0$ the initial state and $m=M, s_M=1$ the final state.
Note that the shift error distribution in $v$ gets rescaled each round: the standard deviation of $G(v)^2_{\Delta, m=0}$ is increasing in each round, but given a $v$, the shift induced in each round in \cref{eq:def_shifted} gets smaller, so that effectively the spread in $p$ stays the same. Thus 
$\Delta_q \approx \Delta$ where $\Delta$ is the initial squeezing.


For the approximate state with $m=M$, i.e. $\ket{V_{\kappa,\mu,M}}$, the mean phase $\theta_p$ is simply the mean $\mu$ of the distribution while the effective squeezing parameter $\Delta_p$ equals
\begin{equation}
\Delta_p = \sqrt{\ln(I_0^2(\kappa)/I_1^2(\kappa))/\pi},
\label{eq:squeeze}
\end{equation}
which for large $\kappa$ becomes $1/\sqrt{\pi \kappa}$, hence directly connecting to the standard deviation of the Gaussian distribution.


Using the formula for linear combinations of trigonometric functions with a phase shift, one can show that the distribution over $u$ of the outgoing state after a round of breeding is again a von Mises distribution (see \eg \cite{book:Mardia.etal.2014:CircularStatistics,book:Jammalamadaka.Sengupta.2001:CircularStatistics} in the context of the convolution of von Mises distributions). Using the convolution property of Gaussian distributions, one can show the same for the $v$ shifts. Combining these two properties, one can show that a round of breeding with measurement outcome $p_{\rm out}$ maps two input states of this form with label $m$ onto an output state of the same form with label $m+1$
\begin{align}
&\ket{V_{\kappa_1,\mu_1, m}}\ket{V_{\kappa_2,\mu_2,  m}} \stackrel{\rm breeding}{\rightarrow} \ket{V_{\kappa_{out},\mu_{out}, m+1}},\nonumber \\
&\kappa_{out}^2 = \kappa_1^2 + \kappa_2^2 + 2\kappa_1 \kappa_2 \cos(\mu_1-\mu_2-2\tilde{p}),\nonumber \\
&\mu_{out} = -\atan2[\kappa_1\cos(\mu_1-\tilde{p}) + \kappa_2\cos(\mu_2+\tilde{p}) \notag\\
 & \hspace{2.35cm},\kappa_1\sin(\mu_1-\tilde{p}) + \kappa_2\sin(\mu_2+\tilde{p})],
\label{eq:breedstep}
\end{align}
with $\tilde{p} = \frac{2\pi}{\xi s_{m+1}}p_{\rm out}$. The details of this derivation can be found in the Appendix \ref{sec:breed}.



Thus, if the two states fed into round $m$ have the error model of \cref{eq:mises}, the outgoing state is of the same type, with new parameters $\kappa_{out}, \mu_{out}$ which depend on measurement outcome $p_{\rm out}$ and the round $m$. Since the ingoing states are normalized, the probability of finding outcome $p_{\rm out}$ can be obtained by evaluating the norm of the outgoing state, see Appendix \ref{sec:breed},
 and we obtain the oscillatory function
\begin{equation}
\mathds{P}(p_{\rm out}) = \frac{\Delta I_0(\kappa_{out}) \mathcal{N}_{out}^2}{\sqrt{\pi} \xi I_0(\kappa_1)I_0(\kappa_2) \mathcal{N}_{1}^2 \mathcal{N}_{2}^2} e^{-\frac{p_{\rm out}^2 \Delta^2}{\xi^2}}.
\label{eq:prob}
\end{equation}
Defining the variable $x = \mu_1-\mu_2-2\tilde{p}\mod 2\pi$ gives a concise description of the effect of one breeding round. The probability $\mathds{P}(x)$ can be simplified in the limit of large initial squeezing, $s_m \Delta \ll 1$ from Eq.~(\ref{eq:prob}). Since $x$ is $2\pi$-periodic, we can use that the limit of a wrapped normal distribution with large variance is simply a circular uniform density of $1/(2\pi)$. Together with the fact that the normalization constants $\mathcal{N}$ all go to $1$ for large initial squeezing, one obtains
\begin{align*}
\kappa_{out}(x) &= \sqrt{\kappa_1^2 +\kappa_2^2 + 2 \kappa_1 \kappa_2 \cos(x)} = (\kappa_1 + \kappa_2)\lambda,\\
\mathds{P}(x) &= \frac{I_0(\kappa_{out})}{2\pi I_0(\kappa_1) I_0(\kappa_2)},
\end{align*}
where we defined $\lambda:= \lambda(x, \kappa_1, \kappa_2)$ with $0\leq \lambda \leq 1$. 


Not surprisingly the growth of $\kappa$ (or shrinking of $\Delta_p$) with the number of rounds is upperbounded as $\kappa_M \leq 2^M \kappa_0$ for any protocol with $M$ rounds and initial states all with equal $\kappa_0$.

To get insight into the probabilistic behavior we would like to bound the probability that $\lambda \leq 1- \epsilon$ for some $\epsilon$ assuming $\kappa_1 \geq 1 / (1-\epsilon)$ and $\kappa_2 \geq 1 / (1-\epsilon)$ in a given round $m$.


Let $\mathds{A} = \{x |\lambda\leq 1-\epsilon\}$, i.e. the set of all events for which $\lambda \leq 1-\epsilon$. Then
\begin{align*}
\mathds{P}(\lambda\leq 1-\epsilon) &= \int_{\mathds{A}} \text{d}x\ \frac{I_0\left((\kappa_1+\kappa_2)\lambda\right)}{2\pi I_0(\kappa_1) I_0(\kappa_2)}, \\
&\leq \frac{I_0\left((\kappa_1+\kappa_2)(1-\epsilon)\right)}{I_0(\kappa_1) I_0(\kappa_2)},
\end{align*}
where we used that $I_0(x)<I_0(y)$ for $x<y$.

It has been shown by Pal'tsev that $\frac{1}{\sqrt{2\pi \kappa}} e^{\kappa-\frac{1}{2\kappa}} \leq I_0(\kappa) \leq \frac{1}{\sqrt{2\pi \kappa}} e^{\kappa+\frac{1}{2\kappa}}$, where the lower bound holds for $\kappa>0$ and the upper bound was only proved for $\kappa>(\sqrt{7}+2)/3$ \cite{Paltsev.1999:BesselBound}.
The range for the upper bound is limited because Pal'tsev derived the bounds for $I_\nu(\kappa)$ with $\nu, \kappa\in \mathds{R}^+_0$. In the special case of $I_0(\kappa)$, it is simple to show that the bound holds for all $\kappa>0$:
$\frac{1}{\sqrt{2\pi \kappa}} e^{\kappa+\frac{1}{2\kappa}}$ is minimal for $\kappa=1$ and $I_0(\kappa), 0\leq \kappa\leq (\sqrt{7}+2)/3$ is maximal for $\kappa=(\sqrt{7}+2)/3$. The bound holds because $\frac{1}{\sqrt{2\pi}} e^{\frac{3}{2}} > I_0((\sqrt{7}+2)/3)$.
Using these bounds, we get
\begin{align}
\mathds{P}(\kappa_{out} \geq (\kappa_1+\kappa_2)(1-\epsilon)) \geq \delta. 
\label{eq:bound_kappa}
\end{align}
 with 
\begin{equation}
\delta\equiv 1 - \sqrt{\frac{2\pi \kappa_1 \kappa_2}{(\kappa_1 + \kappa_2)(1-\epsilon)}} \exp\left(-\epsilon(\kappa_1 + \kappa_2 + 1) +\frac{5}{4}\right).
\label{def:delta}
\end{equation}

For any choice of $\epsilon >0$, this probability is exponentially close to 1 for large $\kappa_1$ or $\kappa_2$.
As a simple example of this bound one can take $\kappa_1=\kappa_2=\kappa_{in}$ and $\epsilon=1/2$. Then we have
\begin{align*}
\mathds{P}(\kappa_{out} \geq \kappa_{in}) \geq 1 - \sqrt{2 \pi \kappa_{in}} &\exp\left(-\kappa_{in} +\frac{3}{4}\right).
\end{align*}
What we see in these bounds is that for sufficiently large $\kappa_{\rm in}$ the protocol produces states with larger $\kappa_{\rm out}$ with high probability.
For example, the probability that $\kappa_{\rm out} \geq \kappa_{\rm in}$ is at least $0.92$ for $\kappa_{\rm in}=5$ (squeezing parameter roughly $\Delta\approx 0.25$). For $\kappa_{\rm in}=10$, the probability that $\kappa_{\rm out} \geq \frac{3}{2} \kappa_{\rm in}$ is at least $0.88$.

Alternatively, one can phrase Eq.~(\ref{eq:bound_kappa}) for large $\kappa$, hence Gaussian-distributed states, in terms of the variance of the Gaussian distribution of shift errors:
In this case, we have that with probability larger than $\delta$ in Eq.~(\ref{def:delta}), the variance of the outgoing state obeys
\begin{align}
& {\rm Var}_{\rm out} \leq \frac{{\rm Var}_{\rm 1,\rm in}{\rm Var}_{2,\rm in}}{(1-\epsilon)({\rm Var}_{1,\rm in}+{\rm Var}_{2,\rm in})}.
\label{eq:vari}
\end{align}
For a grid state with Gaussian distributed shift errors and spacing $\xi$, one has $\Delta_p \approx {\rm Var}/\xi$ so we can see how \cref{eq:vari} expresses the stochastic improvement of the effective squeezing parameter per round.


These bounds are not tight, the probability $\delta$ scales more favorably in practice than these bounds would suggest. In the next section we examine how the mapping of the von Mises distributed states works out numerically as compared to an actual simulation of the protocol with squeezed cat states.

\section{Simulation}
\label{sec:simulation}
\begin{figure}[htb]
\includegraphics[width=\hsize]{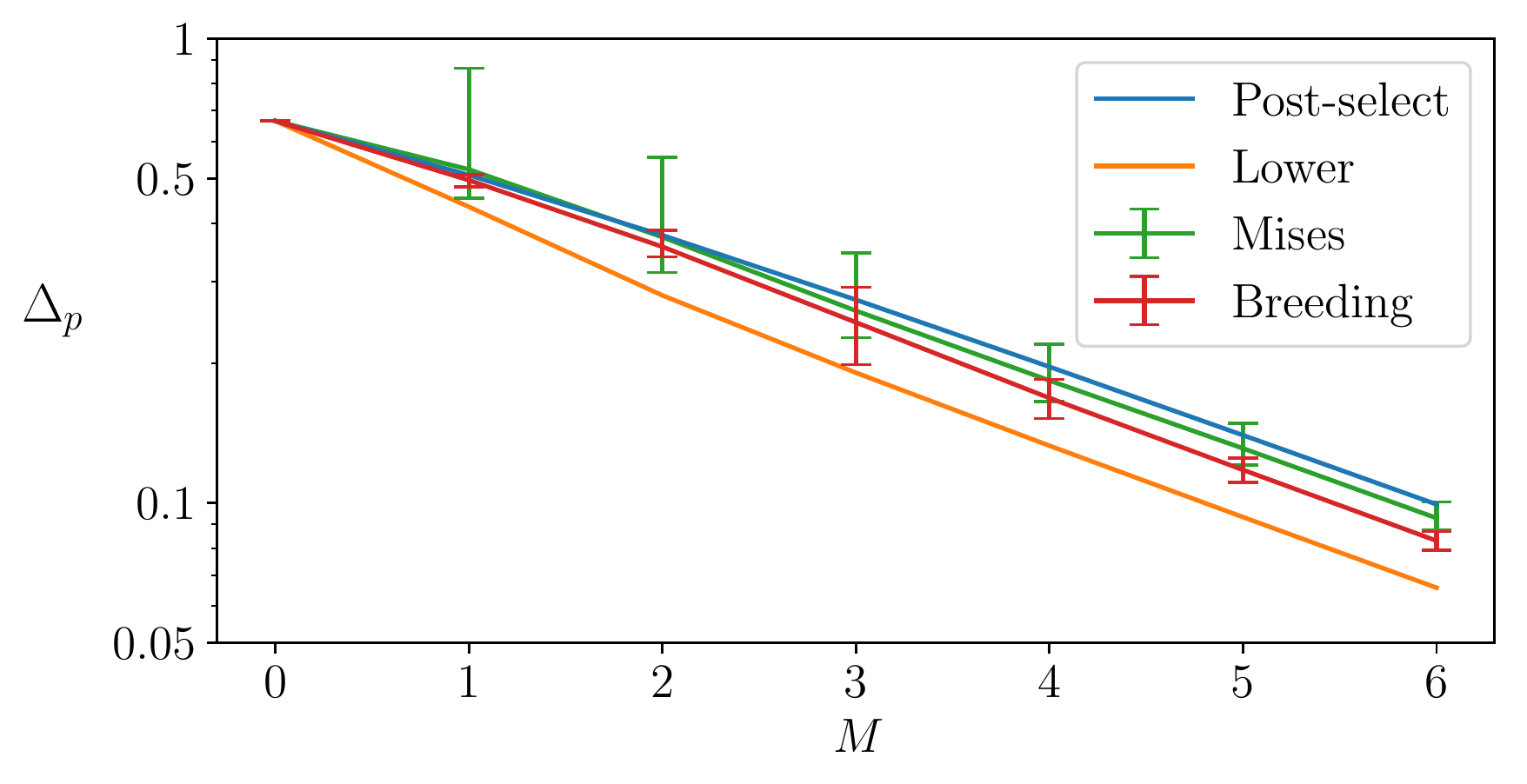}
\caption{Simulated breeding of a sensor state ($S_p = D(\sqrt{\pi})$) with initial squeezing $\Delta=0.2$. Shown is the (dimensionless) effective squeezing parameter $\Delta_p$ (averaged over the homodyne measurement outcomes) versus the number of rounds $M$ of the protocol. `Post-select'  refers to the protocol by Vasconcelos \etal \cite{Vasconcelos.etal.2010:GKPprep}, with squeezed Schr\"odinger cat states as input and post-selected onto the result $p_i=0$ for all measurements. `Breeding' refers to the efficient breeding protocol without post-selection.\ `Mises' is the same efficient breeding protocol, but with von Mises distributed initial states, see \cref{eq:mises}. The error bounds in both Mises and Breeding are asymmetric, i.e. both the variance of all the data above the mean as well as the variance on all the data below the mean are plotted separately. `Lower' is the lower bound for the effective squeezing parameter, namely at round $M$ $\Delta(\kappa_M)=\Delta(2^M \kappa_0)$ where $\Delta(\kappa)$ is given in \cref{eq:squeeze}.
}
\label{fig:simulation}
\end{figure}


 To demonstrate the use of classical post-processing we simulate the breeding of a grid state numerically. 
All the simulated breeding protocols aim to generate an eigenstate of $S_p=D(\sqrt{\pi})$, using $M$ rounds with the efficient breeding protocol.
The breeding is simulated by sampling each measurement result randomly from the state generated by the previous rounds.
This is done for protocols with $M=0,\dots ,6$ rounds, each protocol leading to an approximate grid state with the required spacing $\sqrt{2\pi}$ ($M=0$ means just having a squeezed cat state).

In \cref{fig:simulation} we show the mean and standard deviation of the effective squeezing parameter $\Delta_p$ over 1000 repetitions of this procedure.
In this Figure, the line `Breeding' shows the efficient breeding protocol using finitely-squeezed Schr\"odinger cat states, with $\Delta_q\approx\Delta=0.2$.
This corresponds to states with $\bar{n} \approx 2^M \pi/2 + 25$ photons in all rounds (where 25 is the contribution from initial squeezing by $S(\Delta)$).

In addition, we simulate the same protocol using the von Mises states (with infinite squeezing, corresponding to $\lim_{\Delta\to 0}$ in \cref{eq:mises}) as initial states, starting at a $\kappa$ and $\mu=0$ which gives the same $\Delta_p$ as the squeezed cat states in the real protocol.
For comparison, we also show the effective squeezing achieved by post-selecting onto $p=0$ (`Post-select') and the lower bound (`Lower') on the decrease in the squeezing parameter for the von Mises states as follows from $\kappa_{\rm out} \leq \kappa_1+\kappa_2$.
Since the lower bound has been derived only for von Mises distributed states and not for squeezed cat states as initial states it does not necessarily hold for the latter.
However, it gives a good estimate for the asymptotic behavior, as grid states and the von Mises distributed states get arbitrarily close for small $\Delta_p$.
 
As can be seen the effective squeezing which is achieved on average is lower both for Breeding and Mises than for the post-selected protocol.
Furthermore, the two lines are almost parallel after some rounds, showing that the von Mises error model is a good approximation after a small number of rounds.
All lines show similar scaling with $M$ which we asymptotically expect to be $\sim 2^{-M}$ (this scaling is hard to verify for $M \leq 6$).

\section{Discussion}
\label{sec:discussion}
In this paper we have shown that classical post-processing, combined with the breeding protocol by Vasconcelos \etal yields an efficient method to generate grid states.
By providing a map between breeding and phase estimation, we have argued that any state generated by breeding results in an approximate eigenstate of the commuting displacement operators, i.e. a grid state with an additional known displacement.
We have introduced a new class of approximate grid states which are mapped onto themselves by the application of breeding and allow one to bound the success of the stochastic process implemented by breeding.
In numerical simulations, we could confirm that the protocol discussed in this paper generates grid states reliably, showing scaling close to the asymptotic behavior, even for a small number of rounds.

As we have observed, the action of each round of beam-splitting reduces the spacing of the grid, requiring one to use cat states with large spacing at the beginning of the protocol. An alternative solution is to squeeze the outgoing mode after each beam-splitter so one does not lose a $\sqrt{2}$ factor in each round, see e.g. the use of beam-splitting and $\sqrt{2}$-squeezing in \cite{Glancy.Knill.2006:GKPfaultTolerance}. However, this precisely counteracts the initial squeezing in the $q$-quadrature, hence requires more initial squeezing by $\Delta$. We thus expect that the average number of photons in the initial squeezed cat states scales the same in this alternative protocol, making it a slightly different but not necessarily better alternative.

In any real set-up the measurement of the $p$-quadrature will have some variance, determined for example by the duration of the measurement.
Using the mapping onto phase estimation one can understand this as a spread or uncertainty in the circuit which has been applied to the state, leading to uncertainty of an estimate for the eigenvalue phase.
In the efficient breeding protocol, the spread in $p$ also leads to the preparation of a noisy state which contains additional shift or displacement errors.

While generating optical squeezed Schr\"odinger cat states on demand is a hard task, squeezed cat states with sufficient amplitude to generate the sensor state have been experimentally demonstrated in \cite{Etesse.etal.2015:CatBreedingExp,Huang.etal.2015:CatStatePrep}. The amplitudes of cat states demonstrated there are sufficiently large for 1-2 rounds of breeding with beam splitters only. Multiple rounds could be possibly achieved using additional squeezing in between rounds as suggested in the previous paragraph above.
It might also be of interest to analyze the concrete implementation of this scheme for microwave cavities coupled to superconducting qubits where all components, i.e. the preparation of cat states \cite{Vlastakis.etal.2013:CatStatePrep,Ofek.etal.2016:CatCode}, beam-splitters and homodyne measurement read-out are readily available.
The scheme would lend itself well to a set-up in which cat states are prepared in microwave cavities and are then released \cite{Pfaff.etal.2017:CatchRelease,Yin.etal.2013:CatchRelease} onto transmission lines which couple via beam-splitters and allow for homodyne read-out. 

We would like to thank Kasper Duivenvoorden and Christophe Vuillot for useful discussions and acknowledge support through the EU via the ERC GRANT EQEC. This research was supported in part by Perimeter Institute for Theoretical Physics. Research at Perimeter Institute is supported by the Government of Canada through Industry Canada and by the Province of Ontario through the Ministry of Economic Development $\&$ Innovation.
\bibliography{bibliography}
\appendix
\onecolumngrid

\section{Orthonormality and Completeness of Shifted Grid States}
\label{sec:ortho}
Recall that the shifted grid states are defined as (see \cref{eq:def_shifted}) \begin{align}
\ket{u, v, m}
&=\frac{\sqrt{s_m \xi}}{2\pi}  \sum_{s=-\infty}^{\infty} \exp(i v (s+\frac{u}{2\pi})) \ket{p=s_m \xi 
( s+ \frac{u}{2\pi})},
\label{eq:uvm}
\end{align}
with $u,v\in[-\pi,\pi)$, $s_m\in(0,1]$ and $m\in\mathds{N}_0$. In this section we show that this class of states forms an orthonormal basis. The proof will be split in two parts, showing orthonormality first and completeness afterwards, see the Lemma's below.

If we extend the definition of these basis states so that $u \rightarrow x,v \rightarrow y$ with $x,y \in \mathds{R}$, then we can observe that $\ket{x+2\pi,y,m}=\ket{x,y,m}$ and $\ket{x,y\pm 2\pi,m}=e^{\pm i x} \ket{x,y,m}$. In \cref{sec:breed} we will only consider states of the form 
\begin{align*}
\ket{\Psi} = \int_{-\pi}^\pi \text{d} u \int_{-\pi}^\pi \text{d}v\;  \Theta(u,v)\ket{u,v,m}, 
\end{align*}
where the function $\Theta(x,y)$ is such that $\Theta(x+2\pi,y)=\Theta(x,y)$ and $\Theta(x,y\pm 2\pi)=e^{\mp i x}\Theta(x,y)$. 
For such choice we observe that $\Theta(x,y) \ket{x,y,m}$ is $2\pi$-periodic in both arguments, allowing us to write 
\begin{align}
	\int_{-\pi}^\pi \text{d}x \int_{-\pi}^\pi \text{d}y \ \Theta(x,y)\ket{x,y,m} = \int_{-\pi+z_x}^{\pi+z_x} \text{d}x\ \int_{-\pi+z_y}^{\pi+z_y} \text{d}x\ \Theta(x,y)\ket{x,y,m}.
	\label{eq:shift_int}
	\end{align}
	for any $z_x$ and $z_y$.

\begin{lem}
	\label{lem:uvm_orthonormal}
	The class of shifted grid states as defined in \cref{eq:uvm} is orthonormal, i.\,e.\ it holds that $\braket{u', v', m|u, v, m} = \delta(u-u')\delta(v-v')$.
	\begin{proof}
		From the definition of shifted grid states (\cref{eq:uvm}) and the orthonormality of the momentum eigenstates it follows
		\begin{align*}
		\braket{u', v', m|u, v, m}
		&= \frac{s_m \xi}{(2\pi)^2} \sum_{s, t=-\infty}^\infty \exp\left(i v \left(s+\frac{u}{2\pi}\right) - i v'\left(t+\frac{u'}{2\pi}\right)\right) \delta\left(s_m \xi\left(s-t + \frac{u-u'}{2\pi}\right)\right).
		\end{align*}
		The difference $u-u'$ needs to be an integer multiple of $2\pi$ for the Dirac delta-function to be non-zero. Since $u, u'\in [-\pi,\pi)$, \ie $u-u'\in (-2\pi,2\pi)$, the only solution is $u=u'$ and $s=t$. With $\delta(x) = |a| \delta(a x)$ and $\delta(x) = \frac{1}{2\pi}\sum_{s=-\infty}^\infty \exp(isx)$ the claim follows:
		\begin{align*}
		\braket{u', v', m|u, v, m}
		&= \frac{1}{2\pi}\sum_{s} \exp\left(i (v - v')\left(s+\frac{u}{2\pi} \right)\right) \delta(u-u'),
		\notag \\
		&= \delta(v - v') \delta(u-u').
		\end{align*}
	\end{proof}
\end{lem}
To complete the proof that the shifted grid states form an orthonormal basis, we also show their completeness. We do this by showing $\int\text{d}u\int\text{d}v \ketbra{u,v,m}{u,v,m}p\rangle = \ket{p}$ for any momentum eigenstate $\ket{p}$.
\begin{lem}
	\label{lem:uvm_complete}
	The class of shifted grid states as defined in \cref{eq:uvm} is complete, i.\,e.\ it holds that $\int\text{d}u\int\text{d}v\  \ketbra{u,v,m}{u,v,m} = I$.
	\begin{proof}
		The wave function of a momentum state in the shifted grid state basis is
		\begin{align*}
		\braket{u,v,m|\hat{p}=p}
		&= 
		\sqrt{\frac{s_m \xi}{(2\pi)^2}} \sum_{s=-\infty}^\infty \exp\left(-i v \left(s+\frac{u}{2\pi}\right)\right)
		\delta\left(\xi  s_m \left(s+\frac{u}{2\pi}\right)-p\right).
		\end{align*}
		Since $u\in[-\pi,\pi)$, the Dirac delta distribution is only non-zero for a specific value $s=\tilde{s}$ with $\tilde{p} := p-\xi s_m \tilde{s}, \tilde{p} \in [-\pi,\pi)$. Using $\delta(x) = |a| \delta(a x)$, we can simplify the wave function of a momentum state in the basis of shifted grid states to
		\begin{align*}
		\braket{u,v,m|\hat{p}=p}
		&= 
		\sqrt{\frac{1}{\xi s_m}} \exp\left(-i v \left(\tilde{s}+\frac{u}{2\pi}\right)\right)
		\delta\left(u- \frac{2\pi}{\xi s_m}\tilde{p}\right).
		\end{align*}
		Using the definition of a shifted grid state (see \cref{eq:uvm}) and the wavefunction of a momentum state in the basis of shifted grid states, we obtain
		\begin{align*}
		\iint \text{d}u\ \text{d}v\ \ket{u,v,m}\braket{u,v,m|\hat{p}=p}
		&= 
		\sqrt{\frac{1}{\xi s_m}} \iint \text{d}u\ \text{d}v\ e^{-i v (\tilde{s}+\frac{u}{2\pi}) } 
		\delta\left(u- \frac{2\pi}{\xi s_m}\tilde{p}\right)\ket{u,v,m}\\
		&= 
		\frac{1}{2\pi} \int \text{d}v\ \sum_{s} e^{i v (s-\tilde{s})} 
		\ket{\hat{p}=s_m \xi s+ \tilde{p}} \\
		&= \ket{\hat{p}=s_m \xi \tilde{s}+ \tilde{p}} = \ket{\hat{p}=p}
		\end{align*}
		In the second step, we used the integral representation of the Kronecker delta, $\frac{1}{2\pi} \int_0^{2\pi} \text{d}x \exp\left(ix(n-m)\right) = \delta_{mn}$. 
	\end{proof}
\end{lem}

\section{Analytic Discussion of Breeding}
\label{sec:breed}
In this appendix, we discuss the breeding protocol analytically and show that the class of states used as initial states in \cref{sec:bounds} is closed under the breeding operation. 
To this end, we first analyze the action of breeding on a superposition of shifted grid states $\ket{u,v,m}$ in \cref{sec:breed_sum}, and simplify the state obtained after measurement. Then, we show in \cref{sec:breed_conv} that the action of breeding on the $v$ shifts is that of a convolution of the ingoing wavefunctions, and that a Gaussian error model for these shifts is preserved under breeding. There, we also see that the action on the $u$ shifts is that the ingoing wavefunctions of these shifts are multiplied. Finally in \cref{sec:wavechoice}, we show that for the $u$ shifts, an error model using the von Mises distribution is preserved under breeding, yielding the states used in \cref{sec:bounds}.
\subsection{Breeding Shifted Grid States}
\label{sec:breed_sum}
The action $\mathcal{B}$ of a beam splitter is given by
\begin{align}
&\hat{q}_1\to (\hat{q}_1 - \hat{q}_2)/\sqrt{2}, && \hat{p}_1\to (\hat{p}_1 - \hat{p}_2)/\sqrt{2},\notag \\
&\hat{q}_2 \to (\hat{q}_1 + \hat{q}_2)/\sqrt{2}, && \hat{p}_2\to (\hat{p}_1 + \hat{p}_2)/\sqrt{2}, \label[pluralequation]{eq:bs}
\end{align}
where mode $1$ is the target mode and mode $2$ is the control mode.

Using conjugation one can see that two shifted grid states are transformed as
\begin{align*}
\mathcal{B}\ket{x_1, y_1, m}_1 \ket{x_2, y_2, m}_2 &= \mathcal{B} \frac{s_m \xi}{(2\pi)^2} \sum_{s,t} e^{i (y_2 (s+\frac{x_2}{2\pi})+y_1(t+\frac{x_1}{2\pi}))} e^{i\hat{q}_2 \xi s_m( s+\frac{x_2}{2\pi})}
e^{i\hat{q}_1\xi s_m( t+\frac{x_1}{2\pi})}
\mathcal{B}^\dag \mathcal{B}
\ket{p=0}_1
\ket{p=0}_2
,\notag \\
&=  \frac{s_m \xi}{(2\pi)^2} \sum_{s,t} e^{i (y_2 (s+\frac{x_2}{2\pi})+y_1(t+\frac{x_1}{2\pi}))} e^{i\frac{\hat{q}_1+\hat{q}_2}{\sqrt{2}}\xi s_m( s+\frac{x_2}{2\pi})}
e^{i\frac{\hat{q}_1-\hat{q}_2}{\sqrt{2}}\xi s_m( t+\frac{x_1}{2\pi})}
\ket{p=0}_1
\ket{p=0}_2,\notag \\
&= 
\frac{s_m \xi}{(2\pi)^2} \sum_{s,t} e^{i(y_2 (s+\frac{x_2}{2\pi})+y_1(t+\frac{x_1}{2\pi}))} 
\ket{p=\frac{\xi  s_m}{\sqrt{2}} (t+s+\frac{x_2+x_1}{2\pi})}_1
\ket{p=\frac{\xi  s_m}{\sqrt{2}} (s-t+\frac{x_2-x_1}{2\pi})}_2.
\end{align*}
The invariance of the formal state $\ket{p=0}_1\ket{p=0}_2$ under beam-splitting can be understood from writing $\lim_{\Delta\rightarrow 0} S(1/\Delta) \ket{\rm vac}=\ket{p=0}$ and conjugating the squeezing operators by beam-splitters.

Now, we can easily compute the action of a measurement of mode $2$ with result $p_{\rm out}$:
\begin{align}
& \bra{\hat{p}_2=p_{\rm out}} \mathcal{B}\ket{x_1,y_1,m}_1\ket{x_2,y_2,m}_2=\notag\\
&\hspace{2cm}= 
\frac{s_m \xi}{(2\pi)^2} \sum_{s,t} e^{i (y_2 (s+\frac{x_2}{2\pi})+y_1(t+\frac{x_1}{2\pi})) }
\delta(p_{\rm out}-\frac{\xi  s_m}{\sqrt{2}} (s-t+\frac{x_2-x_1}{2\pi}))
\ket{p=\frac{\xi  s_m}{\sqrt{2}} (t+s+\frac{x_2+x_1}{2\pi})}_1.
\label{eq:measure_uvm}
\end{align}
As a warm-up, we consider the effect of the breeding step on two input modes both in a state of the form 
\begin{align*}
\ket{\Psi_{\text{in}}} = \int_{-\pi}^\pi \text{d}u\ V(u) \ket{u, v, m},
\end{align*}
where $V(u)$ is a wave function with normalization $\int_{-\pi}^{\pi} \text{d}u \ |V(u)|^2=1$ that will be chosen in \cref{sec:wavechoice}.
 Using \cref{eq:measure_uvm}, switching to variables $x$ and $y$, and substituting $\tilde{x}_2 = x_2-\frac{2\pi \sqrt{2}p_{\rm out}}{\xi s_m}$, breeding then gives the output state
\begin{align*}
\ket{\Psi_{\rm out}} &=\frac{s_m \xi}{(2\pi)^2}  \int_{-\pi+z}^{\pi+z} \text{d}\tilde{x}_2
\int_{-\pi}^{\pi} \text{d} x_1\ 
V_1(x_1) V_2(\tilde{x}_2+\frac{2\pi \sqrt{2}p_{\rm out}}{\xi s_m}) 
\sum_{s,t} e^{i (y_2 (s+\frac{\tilde{x}_2}{2\pi}+\frac{\sqrt{2}p_{\rm out}}{\xi s_m})+y_1(t+\frac{x_1}{2\pi})) } \notag \\
&\quad\quad\times
\delta\left(\frac{\xi  s_m}{\sqrt{2}} (s-t+\frac{\tilde{x}_2-x_1}{2\pi})\right)
\ket{p=\frac{\xi  s_m}{\sqrt{2}} (t+s+\frac{\tilde{x}_2+x_1}{2\pi})+p_{\rm out}}_1.
\end{align*}
where $z=\frac{2\pi \sqrt{2} p_{\rm out}}{\xi s_m}$. We can move the integration region for $\tilde{x}_2$ back as described in \cref{eq:shift_int}. Then, note that the Dirac delta distribution is only non-zero if $s=t$ and $\tilde{x}_2=x_1$: After moving the integration region back, $\tilde{x}_2 - x_1\in (-2\pi,2\pi)$. The solutions $\tilde{x}_2 - x_1 = \pm 2\pi$ are a nullset, after applying the Dirac delta distribution, the second integral vanishes for these two solutions. Using $\delta(x) = |a| \delta(a x)$
we obtain:
\begin{align*}
\ket{\Psi_{\rm out}}&= \frac{\sqrt{2}}{2\pi} \int_{-\pi}^\pi\text{d}\tilde{x}_2 \int_{-\pi}^\pi \text{d} x_1\  
V_1(x_1) V_2(\tilde{x}_2+\frac{2\pi \sqrt{2}p_{\rm out}}{\xi s_m}) \notag \\
&\hspace{1cm}\times\sum_{s} e^{i (y_2 (s+\frac{\tilde{x}_2}{2\pi}+\frac{\sqrt{2}p_{\rm out}}{\xi s_m})+y_1(s+\frac{x_1}{2\pi})) } 
\delta(\tilde{x}_2-x_1)
e^{ip_{\rm out}\hat{q}}
\ket{p=\frac{\xi  s_m}{\sqrt{2}} (2s+\frac{\tilde{x}_2+x_1}{2\pi})}_1\\
&= \frac{\sqrt{2}}{2\pi} \int_{-\pi}^\pi \text{d} x_1\ 
V_1(x_1) V_2(x_1+\frac{2\pi \sqrt{2}p_{\rm out}}{\xi s_m}) 
\sum_{s} e^{i ((y_2+y_1) (s+\frac{x_1}{2\pi})+y_2\frac{\sqrt{2}p_{\rm out}}{\xi s_m})}
e^{ip_{\rm out}\hat{q}}
\ket{p=\frac{\xi  s_m}{\sqrt{2}} (2s+\frac{2 x_1}{2\pi})}_1.
\end{align*}
With $s_{m+1} = \sqrt{2}s_m$, we finally have
\begin{align}
\ket{\Psi_{\rm out}}&= 
\frac{\sqrt{2}}{2\pi}  \int_{-\pi}^\pi \text{d} x_1\ 
V_1(x_1) V_2(x_1+\frac{4\pi p_{\rm out}}{\xi s_{m+1}}) 
\sum_{s} e^{i ((y_2+y_1) (s+\frac{x_1}{2\pi})+y_2\frac{2p_{\rm out}}{\xi  s_{m+1}} )}
e^{ip_{\rm out}\hat{q}}
\ket{p=\xi  s_{m+1} (s+\frac{x_1}{2\pi})}_1.
\label{eq:measure}
\end{align}

\subsection{Choice of Wave Function $\Theta(u,v)$}
\label{sec:breed_conv}

We now take the input states in both modes with a wave function $\Theta(x,y)$ (obeying the conditions set forth previously), namely we choose 
\begin{align}
\Theta(u,v)=\frac{1}{\mathcal{N}} V(u) \sum_{s=-\infty}^{\infty}e^{i u s }G_{s_m \Delta}(v+2\pi s),
\label{eq:formtheta}
\end{align}
where $V(u)$ is again the normalized wave function to be chosen in Section \ref{sec:wavechoice}, and $G_{s_m \Delta}$ is a Gaussian distribution
\begin{align*}
G_{\Delta}(v) &= \frac{1}{\sqrt{\Delta\sqrt{\pi }}}\exp(-\frac{v^2}{2\Delta^2}).
\end{align*}
The wave function's dependence on $v$ is thus that of wrapped Gaussian distribution and the $e^{i us}$ factor in \cref{eq:formtheta} is required for the $2\pi$-periodicity of the states as explained below \cref{eq:uvm}. The normalization constant $\mathcal{N}$ is given by
\begin{align}
\mathcal{N}^2 &=  \int_{-\pi}^\pi \text{d} u \int_{-\pi}^\pi \text{d}v\  |V(u)|^2 \sum_{s,t=-\infty}^\infty e^{iu(s-t)} G_{s_m \Delta}(v+2\pi s)G_{s_m \Delta}(v+2\pi t) \notag \\
&=  \int_{-\pi}^\pi \text{d} u \ |V(u)|^2 \sum_{s,t=-\infty}^\infty \frac{1}{2} e^{iu(s-t)} e^{-\frac{\pi^2(s-t)^2}{(s_m \Delta)^2}}\left( \erf\left(\frac{\pi}{s_m \Delta}(s+t+1)\right) - \erf\left(\frac{\pi}{s_m \Delta}(s+t-1)\right)\right) \xrightarrow{s_m\Delta \ll 1} 1.
\label{eq:norm}
\end{align}
In the limit $s_m \Delta \ll 1$, the exponential $e^{-\frac{\pi^2(s-t)^2}{(s_m \Delta)^2}}$ enforces $s-t=0$, while for the difference of error functions to be non-zero, we need $s+t=0$, hence together one has $s=t=0$. Note that $s_m\in(0,1]$, \ie if $\Delta\ll 1$, then also $s_m\Delta\ll 1$.

Using this wave function we can write the input state in one of the modes as
\begin{align*}
\ket{\Psi_{\rm in}} = \frac{1}{\mathcal{N}} \int_{-\pi}^\pi \text{d} u \int_{-\pi}^\pi \text{d}v\  V(u) \sum_{s=-\infty}^\infty e^{ius} G_{s_m \Delta}(v+2\pi s) \ket{u, v, m} = \frac{1}{\mathcal{N}} \int_{-\pi}^\pi \text{d} x \int_{-\infty}^\infty \text{d}y\ V(x)G_{s_m \Delta}(y) \ket{x, y, m}.
\end{align*}

We will assume that the Gaussian wave function of both modes has the same variance, and mean equal to $0$. This choice is justified if the outgoing Gaussians only depend on the round $m$, which we will show below. 

From the result for breeding states with arbitrary superpositions of shifts in $\hat{p}$,  \cref{eq:measure}, it follows that
\begin{align*}
\ket{\Psi_{\rm out}}&= 
\frac{\sqrt{2}}{2\pi \mathcal{N}^2} \int_{-\pi}^\pi \text{d} x_1 \int_{-\infty}^\infty\text{d}y_1
\int_{-\infty}^\infty\text{d}y_2\ 
V_1(x_1) V_2(x_1+\frac{4\pi p_{\rm out}}{\xi s_{m+1}})G_{s_m\Delta}(y_1)G_{s_m\Delta}(y_2)  \notag \\
&\quad\quad\times \sum_{s} e^{i ((y_1+y_2) (s+\frac{x_1}{2\pi})+y_2\frac{2p_{\rm out}}{\xi  s_{m+1}} )}e^{ip_{\rm out}\hat{q}}
\ket{p=\xi  s_{m+1} (s+\frac{x_1}{2\pi})},\\
&= 
\frac{\sqrt{2}}{2\pi \mathcal{N}^2} \int_{-\pi}^\pi \text{d} x_1 \int_{-\infty}^\infty\text{d}y_1
\int_{-\infty}^\infty\text{d}y_2\ 
V_1(x_1) V_2(x_1+\frac{4\pi p_{\rm out}}{\xi s_{m+1}}) G_{s_m\Delta}(\tilde{y}-y_2) G_{s_m\Delta}(y_2) \notag \\
&\quad\quad\times \sum_{s} e^{i (\tilde{y} (s+\frac{x_1}{2\pi})+y_2\frac{2p_{\rm out}}{\xi  s_{m+1}} )}e^{ip_{\rm out}\hat{q}}
\ket{p=\xi  s_{m+1} (s+\frac{x_1}{2\pi})}\\
&= 
\frac{\sqrt{2s_{m+1}\Delta\sqrt{\pi}}}{2\pi \mathcal{N}^2} e^{-\frac{p_{\rm out}^2\Delta^2}{2\xi^2 }} \int_{-\pi}^\pi \text{d} x_1 \int_{-\infty}^\infty\text{d}\tilde{y}\ 
V_1(x_1) V_2(x_1+\frac{4\pi p_{\rm out}}{\xi s_{m+1}}) G_{s_{m+1}\Delta}(\tilde{y}) \notag \\
&\quad\quad\times \sum_{s} e^{i (\tilde{y} (s+\frac{x_1}{2\pi})+\tilde{y}\frac{p_{\rm out}}{\xi  s_{m+1}} )}
e^{ip_{\rm out}\hat{q}}\ket{p=\xi  s_{m+1} (s+\frac{x_1}{2\pi})}.
\end{align*}

Here, we used  $s_{m+1}=\sqrt{2}s_m$ and the substitution $\tilde{y}=y_1+y_2$ to write the integral over $y_2$ as a {\em convolution} of Gaussian wave functions.
Comparing this state with the definition of shifted grid states, \cref{eq:uvm}, we see that the outgoing state has a `simple' expression in terms of the shifted grid states with extended parameters:
\begin{align*}
\ket{\Psi_{\rm out}}
&= 
\frac{\sqrt{2\Delta\sqrt{\pi}}}{\sqrt{\xi} N^2} e^{-(\frac{p_{\rm out}\Delta}{\xi \sqrt{2}})^2} \int_{-\pi}^\pi \text{d} x_1 \int_{-\infty}^\infty\text{d}\tilde{y}\ 
V_1(x_1) V_2(x_1+\frac{4\pi p_{\rm out}}{\xi s_{m+1}})G_{s_{m+1}\Delta}(\tilde{y})
\ket{x_1+\frac{2 \pi p_{\rm out}}{\xi  s_{m+1}}, \tilde{y},m+1}.
\end{align*}
With $\tilde{x} = x_1+\frac{ 2\pi p_{\rm out}}{\xi  s_{m+1}}$ 
and \cref{eq:shift_int}, we finally have
\begin{align}
\ket{\Psi_{out}}
&= 
\frac{\sqrt{2\Delta\sqrt{\pi}}}{\sqrt{\xi} \mathcal{N}^2} e^{-(\frac{p_{\rm out}\Delta}{\xi \sqrt{2}})^2} \int_{-\pi}^\pi \text{d} \tilde{x} \int_{-\infty}^\infty\text{d}\tilde{y}\ 
V_1(\tilde{x}-\frac{2\pi p_{\rm out}}{\xi s_{m+1}})V_2(\tilde{x}+\frac{2\pi p_{\rm out}}{\xi s_{m+1}})G_{s_{m+1}\Delta}(\tilde{y})
\ket{\tilde{x},\tilde{y},m+1} \label{eq:v_out_xy}\\
&= 
\frac{\sqrt{2\Delta\sqrt{\pi}}}{\sqrt{\xi} \mathcal{N}^2} e^{-(\frac{p_{\rm out}\Delta}{\xi \sqrt{2}})^2} \int_{-\pi}^\pi \text{d} u \int_{-\pi}^\pi\text{d}v\ 
V_1(u-\frac{2\pi p_{\rm out}}{\xi s_{m+1}})V_2(u+\frac{2\pi p_{\rm out}}{\xi s_{m+1}})\sum_{s=-\infty}^\infty e^{ius} G_{s_{m+1}\Delta}(v+2\pi s)
\ket{u,v,m+1}
\label{eq:v_out_uv}
\end{align}
Hence we conclude that the outgoing state has the same wave function dependence in $v$ as the ingoing states. The only change is $s_m \to s_{m+1}$. From this last equation we can also immediately see the action of breeding on the wave function $V(u)$, i.e. $V(u) \rightarrow  V(u+\frac{2\pi p_{\rm out}}{\xi s_{m+1}})V(u-\frac{2\pi p_{\rm out}}{\xi s_{m+1}})$.


\subsection{Choice for Wave Function $V(u)$}
\label{sec:wavechoice}
As can be seen in \cref{eq:v_out_uv}, the output state depends on a product of the form $V_1(u)V_2(u)$. For some choices for the ingoing wavefunctions, one can simplify $V_1(u)V_2(u) = V_{\text{out}}(u)$, where all $V_i$ are in the same class of functions. One such class of functions is the set of von Mises distributions, which is closed under multiplication. Let 
\begin{align}
V(x-\mu)_{\kappa} = \frac{\exp\left(\frac{\kappa}{2}\cos(x-\mu)\right)}{\sqrt{2\pi I_0(\kappa)}}.
\label{eq:vonMises}
\end{align}

Assuming a von Mises wave function in $u$ and a wrapped (signed) Gaussian wave function in $v$, the initial state of the system is thus chosen as 
\begin{align}
\ket{\Psi_{\rm in}} = \frac{1}{\mathcal{N}} \int_{-\pi}^\pi \text{d} u \int_{-\pi}^\pi \text{d}v\  V_\kappa(u-\mu) \sum_{s=-\infty}^\infty e^{ius} G_{s_m \Delta}(v+2\pi s) \ket{u, v, m} = \frac{1}{\mathcal{N}} \int_{-\pi}^\pi \text{d} x \int_{-\infty}^\infty \text{d}y\ V_\kappa(x-\mu) G_{s_m \Delta}(y) \ket{x, y, m},
\label{eq:u_in}
\end{align}
where $V_{\kappa}(u)$ is the distribution defined in \cref{eq:vonMises}. The normalization constant $\mathcal{N}$ has the same form as \cref{eq:norm}, with the von Mises wave function defined above taking the role of $V(x)$. This is also the initial state used in the main text, see \cref{eq:mises}. Using the result for a Gaussian error model in $\hat{q}$ and an arbitrary wave function for $\hat{p}$,  \cref{eq:v_out_xy}, the state after measurement is
\begin{align}
\ket{\Psi_{\rm out}}
&= 
\frac{\sqrt{2\Delta\sqrt{\pi}}}{\sqrt{\xi} \mathcal{N}_2 \mathcal{N}_1} e^{-(\frac{p_{\rm out}\Delta}{\xi \sqrt{2}})^2} \int_{-\pi}^\pi \text{d} \tilde{x} \int_{-\infty}^\infty\text{d}\tilde{y}\ 
V_{\kappa_2}(\tilde{x}+\frac{2\pi p_{\rm out}}{\xi s_{m+1}}-\mu_2)V_{\kappa_1}(\tilde{x}-\frac{2\pi p_{\rm out}}{\xi s_{m+1}}-\mu_1)G_{s_{m+1}\Delta}(\tilde{y})
\ket{\tilde{x},\tilde{y},m+1},
\end{align}
where $\mathcal{N}_1, \mathcal{N}_2$ are the normalization constants of the initial state of modes $1$ and $2$, respectively. This expression can be simplified with the following lemma.
\begin{lem}
	For a product of von Mises wave functions as defined in \cref{eq:vonMises} it holds that
	\begin{align*}
	V_{\kappa_1}(x-\mu_1)V_{\kappa_2}(x-\mu_2) &= \sqrt{\frac{I_0(\kappa)}{2\pi I_0(\kappa_1) I_0(\kappa_2)}} V_{\kappa}(x-\mu),
	\end{align*}
	with
	\begin{align*}
	&\mu = -\atan2\left(\kappa_1\cos(\mu_1) + \kappa_2\cos(\mu_2), \kappa_1\sin(\mu_1) + \kappa_2\sin(\mu_2)\right),
	&&\kappa^2 = \kappa_1^2+\kappa_2^2+2\kappa_1\kappa_2\cos(\mu_1-\mu_2).
	\end{align*}
	\begin{proof}
		We can use the properties of linear combinations of trigonometric functions to show that the set of von Mises distributions is closed under multiplication. We have
		\begin{align*}
		V(x-\mu_1)_{\kappa_1}V(x-\mu_2)_{\kappa_2} &= \frac{\exp\left(\frac{\kappa_1}{2}\cos(x-\mu_1) + \frac{\kappa_2}{2}\cos(x-\mu_2)\right)}{2\pi \sqrt{I_0(\kappa_1)I_0(\kappa_2)}}
		\end{align*}
		For the exponent on the r.\,h.\,s.\ it holds that
		\begin{align*}
		\kappa_1\cos(x-\mu_1) + \kappa_2\cos(x-\mu_2) &=
		\left(\kappa_1\cos(\mu_1) + \kappa_2 \cos(\mu_2)\right)\cos(x)
		+ \left(\kappa_1\sin(\mu_1) + \kappa_2\sin(\mu_2)\right)\sin(x)\\
		&=\sqrt{\kappa_1^2+\kappa_2^2+2\kappa_1\kappa_2\cos(\mu_1-\mu_2)} \cos(x-\mu) := \kappa \cos(x-\mu) 
		\end{align*}
		with $\mu, \kappa$ as in the claim.
		In the first step, we used $\cos(x-y) = \cos(x)\cos(y) + \sin(x)\sin(y)$. In the second step, we used $a\cos(x) + b\sin(x) = \sqrt{a^2+b^2}\cos(x+\atan2(a, b))$.
	\end{proof}
\end{lem}

Using this lemma, the outgoing state is given by
\begin{align*}
\ket{\Psi_{\rm out}}
&= 
\sqrt{\frac{I_0(\kappa) \Delta}{\sqrt{\pi} I_0(\kappa_1) I_0(\kappa_2) \xi \mathcal{N}_2^2 \mathcal{N}_1^2}} e^{-\frac{p_{\rm out}^2\Delta^2}{2\xi^2}} \int_{-\pi}^\pi \text{d} u \int_{-\pi}^\pi\text{d}v\ 
V_{\kappa}(u-\mu)\sum_{s=-\infty}^\infty e^{ius} G_{s_{m+1}\Delta}(v+2\pi s),
\ket{u,v,m+1}
\end{align*}
with
\begin{align*}
\mu &= -\atan2\left(\kappa_2\cos(\mu_2 - \frac{2\pi p_{\rm out}}{\xi s_{m+1}}) + \kappa_1\cos(\mu_1 + \frac{2\pi p_{\rm out}}{\xi s_{m+1}}), \kappa_2\sin(\mu_2 - \frac{2\pi p_{\rm out}}{\xi s_{m+1}}) + \kappa_1\sin(\mu_1 + \frac{2\pi p_{\rm out}}{\xi s_{m+1}})\right)\\
\kappa^2 &= \kappa_2^2+\kappa_1^2+2\kappa_2\kappa_1\cos(\mu_2-\mu_1-\frac{4\pi p_{\rm out}}{\xi s_{m+1}})
\end{align*}
This state is not yet normalized. However, we can use \cref{eq:norm} to obtain $\mathcal{N}_{out}$ and both normalize this state and obtain the probability distribution of measurement results $p_{\rm out}$ as written in the main text.

\end{document}